\documentclass[]{pasj01}

\begin{document} 
\Received{}
\Accepted{}

\title{CHORUS. I. Cosmic HydrOgen Reionization Unveiled with Subaru: Overview}

\author{Akio K.\ \textsc{Inoue}\altaffilmark{1,2,3}%
}\email{akinoue@aoni.waseda.jp}
\author{Satoshi \textsc{Yamanaka}\altaffilmark{2,3}}
\author{Masami \textsc{Ouchi}\altaffilmark{4,5,6}}
\author{Ikuru \textsc{Iwata}\altaffilmark{4,7}}
\author{Kazuhiro \textsc{Shimasaku}\altaffilmark{8,9}}
\author{Yoshiaki \textsc{Taniguchi}\altaffilmark{10,11}}
\author{Tohru \textsc{Nagao}\altaffilmark{11}}
\author{Nobunari \textsc{Kashikawa}\altaffilmark{8}}
\author{Yoshiaki \textsc{Ono}\altaffilmark{5}}
\author{Ken \textsc{Mawatari}\altaffilmark{5,3}}
\author{Takatoshi \textsc{Shibuya}\altaffilmark{12}}
\author{Masao \textsc{Hayashi}\altaffilmark{4}}
\author{Hiroyuki \textsc{Ikeda}\altaffilmark{13}}
\author{Haibin \textsc{Zhang}\altaffilmark{14,5}}
\author{Yongming \textsc{Liang}\altaffilmark{7,4,8}}
\author{C.-H. \textsc{Lee}\altaffilmark{15}}
\author{Miftahul \textsc{Hilmi}\altaffilmark{5,8}}
\author{Satoshi \textsc{Kikuta}\altaffilmark{16}}
\author{Haruka \textsc{Kusakabe}\altaffilmark{17}}
\author{Hisanori \textsc{Furusawa}\altaffilmark{4,7}}
\author{Tomoki \textsc{Hayashino}\altaffilmark{18}}
\author{Masaru \textsc{Kajisawa}\altaffilmark{11}}
\author{Yuichi \textsc{Matsuda}\altaffilmark{4,7}}
\author{Kimihiko \textsc{Nakajima}\altaffilmark{4}}
\author{Rieko \textsc{Momose}\altaffilmark{8}}
\author{Yuichi \textsc{Harikane}\altaffilmark{19,4}}
\author{Tomoki \textsc{Saito}\altaffilmark{20}}
\author{Tadayuki \textsc{Kodama}\altaffilmark{21}}
\author{Shotaro \textsc{Kikuchihara}\altaffilmark{5,8}}
\author{Masanori \textsc{Iye}\altaffilmark{4}}
\author{Tomotsugu \textsc{Goto}\altaffilmark{22}}

\altaffiltext{1}{Department of Physics, School of Advanced Science and Engineering, Faculty of Science and Engineering, Waseda University, 3-4-1, Okubo, Shinjuku, Tokyo 169-8555}

\altaffiltext{2}{Waseda Research Institute for Science and Engineering, Faculty of Science and Engineering, Waseda University, 3-4-1, Okubo, Shinjuku, Tokyo 169-8555}

\altaffiltext{3}{Department of Environmental Science and Technology, 
Faculty of Design Technology, Osaka Sangyo University,
3-1-1 Nakagaito, Daito, Osaka 574-8530}

\altaffiltext{4}{National Astronomical Observatory of Japan, 
2-21-1, Osawa, Mitaka, Tokyo, 181-8588}

\altaffiltext{5}{Institute for Cosmic Ray Research, The University of
Tokyo, 5-1-5 Kashiwa-no-ha, Kashiwa, Chiba 277-8582}

\altaffiltext{6}{Kavli Institute for the Physics and Mathematics of the
Universe (Kavli IPMU), WPI, The University of Tokyo, Kashiwa, Chiba
277-8583}

\altaffiltext{7}{Department of Astronomical Science, The Graduate University for Advanced Studies, SOKENDAI, 2-21-1, Osawa, Mitaka, Tokyo 181-8588}

\altaffiltext{8}{Department of Astronomy, Graduate School of Science, The University of Tokyo, 7-3-1 Hongo, Bunkyo-ku, Tokyo 113-0033}

\altaffiltext{9}{Research Center for the Early Universe, Graduate School of Science, The University of Tokyo, 7-3-1 Hongo, Bunkyo-ku, Tokyo 113-0033}

\altaffiltext{10}{The Open University of Japan, 2-11 Wakaba, Mihama-ku, Chiba, Chiba 261-8586}

\altaffiltext{11}{Research Center for Space and Cosmic Evolution, 
Ehime University, Matsuyama, Ehime 790-8577}

\altaffiltext{12}{Kitami Institute of Technology, 165 Koen-cho, Kitami, Hokkaido 090-8507}

\altaffiltext{13}{National Institute of Technology, Wakayama College, Gobo, Wakayama 644-0023}

\altaffiltext{14}{Department of Astronomy, Tsinghua University, No.1 Qinghuayuan, Beijing 100084, China}

\altaffiltext{15}{NSF's National Optical-Infrared Astronomy Research Laboratory, Tucson, AZ 85719, USA}

\altaffiltext{16}{Center for Computational Sciences, University of Tsukuba, 1-1-1 Tennodai, Tsukuba, Ibaraki 305-8577}

\altaffiltext{17}{Observatoire de Gen\`{e}ve, Universit\'e de Gen\`{e}ve, 51 chemin de P\'egase, 1290 Versoix, Switzerland}

\altaffiltext{18}{Research Center for Neutrino Science, Tohoku University, Aramaki, Aoba-ku,
Sendai 980-8578}

\altaffiltext{19}{Department of Physics and Astronomy, University College London, Gower Street, London WC1E 6BT, UK}

\altaffiltext{20}{Nishi-Harima Astronomical Observatory, Centre for Astronomy, University of Hyogo, 407-2 Nishigaichi, Sayo, Sayo-gun, Hyogo 679-5313} 

\altaffiltext{21}{Astronomical Institute, Tohoku University, 6-3 Aramaki, Aoba-ku, Sendai 980-8578} 

\altaffiltext{22}{Institute of Astronomy, National Tsing Hua University, 101 Section 2 Kuang Fu Road, Hsinchu 300, Taiwan, R. O. C.}



\maketitle

\begin{abstract}
To determine the dominant sources for cosmic reionization, the evolution history of the global ionizing fraction, and the topology of the ionized regions, we have conducted a deep imaging survey using four narrow-band (NB) and one intermediate-band (IB) filters on the Subaru/Hyper Suprime-Cam (HSC), called Cosmic HydrOgen Reionization Unveiled with Subaru (CHORUS).
The central wavelengths and full-widths-at-half-maximum of the CHORUS filters are, respectively, 386.2 nm and 5.5 nm for NB387, 526.0 nm and 7.9 nm for NB527, 717.1 nm and 11.1 nm for NB718, 946.2 nm and 33.0 nm for IB945, and 971.2 nm and 11.2 nm for NB973.
This combination, including NB921 (921.5 nm and 13.5 nm) from the Subaru Strategic Program with HSC (HSC SSP), are carefully designed, as if they were playing a chorus, to observe multiple spectral features simultaneously, such as Lyman continuum, Ly$\alpha$, C~{\sc iv}, and He~{\sc ii} for $z=2$--$7$.
The observing field is the same as that of the deepest footprint of the HSC SSP in the COSMOS field and its effective area is about 1.6 deg$^2$.
Here, we present an overview of the CHORUS project, which includes descriptions of  the filter design philosophy, observations and data reduction, multiband photometric catalogs, assessments of the imaging quality, measurements of the number counts, and example use cases of the data.
All the imaging data, photometric catalogs, masked pixel images, data of limiting magnitudes and point spread functions, results of completeness simulations, and source number counts are publicly available through the HSC SSP database.
\end{abstract}

\section{Introduction}


Understanding cosmic reionization is one of the most important objectives in observational cosmology. 
Measurements of the Thomson scattering optical depth of free electrons in the intergalactic medium (IGM) in the cosmic microwave background with {\it Wilkinson Microwave Anisotropy Probe} and {\it Planck} satellites place the epoch of hydrogen reionization at $z\sim10$ \citep{Komatsu2011,Planck2018}. 
Detection of the Gunn-Peterson trough in the spectra of quasars at $z>6$ (e.g., \cite{Fan2006,Becker2007}) and the decrement in the luminosity functions (LFs) of Ly$\alpha$ emitters (LAEs) at $z>6$ (e.g., \cite{Kashikawa2006,Ouchi2010,Hu2010,Kashikawa2011,Nakamura2011,Konno2014,Matthee2015,Santos2016,Konno2018,Itoh2018,Hu2019,Taylor2020}) suggest a rapid increase in the hydrogen neutral fraction, $x_{\rm HI}$, beyond $z=6$. 
The $x_{\rm HI}$ measurements obtained so far still has a large scatter due to the small statistics and the systematic uncertainty in each measurement (e.g., \cite{Robertson2015,GreigMesinger17,Inoue2018,Finkelstein2019,Naidu2020}).
New $x_{\rm HI}$ measurements near $z\sim7$ based on large samples of galaxies over large cosmic volumes are required to reach a concordance in the history of reionization.

Galaxies, such as LAEs and Lyman break galaxies (LBGs), at $z>6$ can be the dominant ionizing sources if their Lyman continuum (LyC) emissivity or escape fraction $f_{\rm esc}$, is sufficiently high. 
The required $f_{\rm esc}$ is $\sim20\%$, assuming a standard population synthesis model (e.g., \cite{Inoue2006,Robertson2015,Finkelstein2019,Naidu2020}). 
Measuring the LyC emissivity of these galaxies, which is still quite uncertain, is the most critical step in the determination of the ionization sources for reionization. 
Bright quasars do not contribute much to the LyC emissivity for reionization owing to their rarity (e.g., \cite{Bianchi2001}), whereas faint Active Galactic Nuclei (AGNs) might contribute to or even dominate \citep{Giallongo2015,Madau2015}. 
However, other observations suggest that faint AGNs are not sufficiently abundant \citep{Kashikawa2015,Onoue2017,Matsuoka2018,Parsa2018}. 
Measuring the LyC emissivity of AGNs is also required to resolve their role in reionization \citep{Micheva2017,Grazian2018}. 
The metal-free stellar population, the so-called Population III (Pop III) stars (e.g., \cite{Carlberg1981}), can be an important ionizing source, particularly in the early phase of the reionization process (e.g., \cite{Sokasian2004}).
Because their characteristics, such as the initial mass function (IMF), are unknown observationally owing to the lack of information \citep{Nagao2005,Nagao2008,Kashikawa2012,Vanzella2020}, identifying Pop~III clusters and revealing their nature are highly relevant to understanding their role in reionization.

Depending on the dominant ionizing source, the topology of reionization is expected to be different: ``inside-out'' \citep{Iliev2006} by galaxies or ``outside-in'' \citep{Miralda-Escude2000} by X-ray sources, such as AGNs. 
Observationally, the spatial inhomogeneity of $x_{\rm HI}$ at $z\sim6$ has been reported to be based on quasar spectra \citep{Becker2015} and LAE LFs \citep{Ouchi2010,Nakamura2011}. 
Obtaining the overdensities of LAEs at $z\gtrsim6.5$ may support the concept that these LAEs are located in the ionized bubbles produced by the galaxy overdensities (e.g., \cite{Castellano2016,Bagley2017,Higuchi2019,Harikane2019,Tilvi2020}).
Resolving the ionization topology is one of the main scientific goals for future H~{\sc i} 21~cm tomography with the Square Kilometer Array (SKA) (e.g., \cite{Hasegawa2016}). 
However, with a sufficiently large survey of LAEs and LBGs, we may visualize the $x_{\rm HI}$ map before the SKA era. 
Such topology observations will be important consistency checks of the ionizing source observations, by comparing the observed topology with the prediction based on the identified dominant ionizing sources.

The Hyper Suprime-Cam (HSC) \citep{Miyazaki2018,Komiyama2018,Kawanomoto2018,Furusawa2018} has the {\it widest} field-of-view (FoV), 1.75 deg$^2$, on an 8m-class telescope. 
This unique capability with a well-considered set of narrowband (NB) filters, for the first time, allows dealing with the central questions of reionization by an unprecedented wide and deep NB survey. 
Since 2007, we have studied the specifications of HSC NBs carefully (see section~2) and developed them under strict scientific reviews with the financial support of approximately 100 M yen from the Japan Society for the Promotion of Science. 
Four NBs (NB387, NB816, NB921, and NB101) among them are included in the on-going HSC Subaru Strategic Program (SSP) \citep{Aihara2018a} for deep surveys of LAEs at $z\simeq2.2$, 5.7, 6.6, and 7.3 \citep{Ouchi2018}. 
The latter three high-$z$ surveys provide new estimates of the cosmic average of $x_{\rm HI}$ at $z\simeq6.6$ and 7.3 from comparison with LAE LFs \citep{Ouchi2018,Shibuya2018a,Shibuya2018b,Konno2018,Inoue2018}. 
However, the HSC SSP cannot address the LyC emissivity and the $x_{\rm HI}$ topology directly.

Thus, we initiated a Subaru intensive program, Cosmic HydrOgen Reionization Unveiled with Subaru (CHORUS), with NB387, NB527, NB718, IB945, and NB973 (Table~\ref{tab:filters}; see Fig.~\ref{fig:filters}). 
Supplying the data of the NB921 and broadband (BB) filters from the HSC SSP, we aim to answer the questions of the ionization source and the topology, in addition to the history, by measuring the LyC emissivity values of galaxies and AGNs, abundance of Pop~III stars, and LF of faint AGNs, and visualizing the $x_{\rm HI}$ map directly see sections 6.1-6.5 for more information). 
These measurements are realized only in combination with our NBs.
This program will become a legacy survey at least for a decade because no other deep and wide-area survey, such as the Large Synoptic Survey Telescope, has a set of NBs like this program.


This paper presents an overview of the CHORUS project and its first public data release: CHORUS PDR1.
In section~2, we describe the CHORUS filter set in detail.
Section~3 is a summary of the observations, data reduction, and photometric catalogs of the CHORUS data.
Section~4 presents a summary of imaging quality, such as the size of the point spread function and the limiting magnitude of each NB.
In section~5, the number counts in the five CHORUS NBs are given as a set of fundamental measurements of the NB imaging data.
Section~6 gives a brief instruction of the usage of the CHORUS PDR1 data and some example science cases.
A summary of this paper is found in the final section.

We assume a standard set of cosmological parameters of $H_0=70$ km s$^{-1}$ Mpc$^{-1}$, $\Omega_M=0.3$ and $\Omega_\Lambda=0.7$ and use the AB magnitude system \citep{Oke1990} throughout the study.

\section{Filter set}

\begin{figure}
  \begin{center}
    \includegraphics[width=70mm]{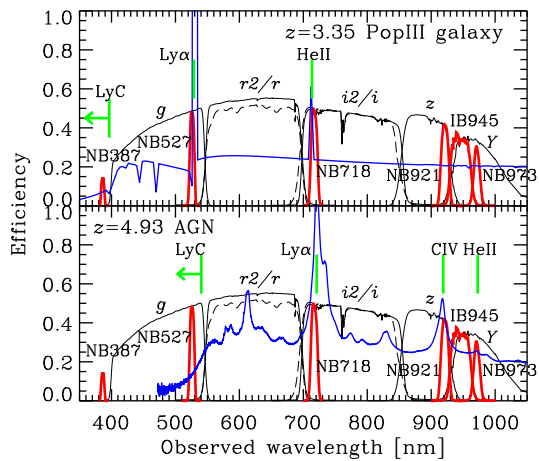}
  \end{center}
  \caption{Filter efficiency functions (narrowbands and broadbands are shown in red and black, respectively) with example spectra of a galaxy at $z=3.35$ (top) and an AGN at $z=4.93$ (bottom). For the $r2/r$ and $i2/i$ bands, the solid lines show the $r2$ and $i2$ transmissions and the dashed lines show the $r$ and $i$ transmissions. The effects considered in addition to the filter transmission are as follows: atmospheric transmission at the summit of Mauna Kea (airmass of 1.2 and water vapor column of 1.6 mm), the primary mirror reflectance, and the instrumental throughput of the HSC including the CCD quantum efficiency and an average vignetting effect in the field-of-view.} \label{fig:filters}
\end{figure}

\begin{table*}
  \tbl{Summary of filter properties.}{%
  \begin{tabular}{llllllllll}
      \hline
      & \multicolumn{3}{l}{Area-weighted mean (Central position)} 
      & & \multicolumn{4}{l}{Redshift range of targeted features$^d$} \\
      \cline{2-4} \cline{6-9}
      Filter & CW$^a$ & FWHM$^b$ & TP$^c$ &  & LyC & Ly$\alpha$ & C~{\sc iv} & He~{\sc ii} & Papers which used the NBs.$^e$ \\ 
      \hline
      NB387 & 386.2 (385.8) & 5.5 (5.7) & 0.842 (0.871) & & $>$3.27 & 2.15--2.20 & --- & --- &  A18a, L20 \\
      NB527 & 526.0 (525.5) & 7.9 (7.5) & 0.956 (0.922) & & $>$4.81 & 3.29--3.36 & --- & 2.18--2.23 & H20 \\
      NB718 & 717.1 (716.8) & 11.1 (11.1) & 0.927 (0.911) & & --- & 4.85--4.94 & --- & 3.34--3.41 & Z20, H20 \\
      NB921$^f$ & 921.5 (920.4) & 13.5 (13.4) & 0.935 (0.937) & & --- & 6.52--6.64 & 4.90--4.99 & --- & A18a, O18, H20 \\
      IB945 & 946.2 (944.7) & 33.0 (32.2) & 0.955 (0.952) & & --- & 6.65--6.92 & --- & --- & \\
      NB973 & 971.2 (969.9) & 11.2 (11.0) & 0.878 (0.834) & & --- & 6.94--7.04 & --- & 4.88--4.95 & I18, H20 \\
      \hline
    \end{tabular}}\label{tab:filters}
\begin{tabnote}
$^a$ The central wavelength (CW) of the transmission function in units of nm.\\
$^b$ The full-width-at-half-maximum (FWHM) of the transmission function in units of nm.\\
$^c$ The transmission peak (TP) of the transmission function itself (i.e., not including the instrumental and atmospheric throughput).\\
$^d$ (CW$\pm$FWHM/2)/$\lambda$-1. Adopted rest-frame wavelength ($\lambda$) of the features: Lyman limit: 91.175 nm, Ly$\alpha$: 121.567 nm, C~{\sc iv}: 154.949 nm, and He~{\sc ii}: 164.033 nm. \\
$^e$ References: A18a: Aihara et al.~(2018a), L20: Liang et al.~(2020), 
H20: Hayashi et al.~(2020), Z20: Zhang et al.~(2020), O18: Ouchi et al.~(2018), 
I18: Itoh et al.~(2018)\\
$^f$ The imaging data collected and presented in SILVERRUSH (Ouchi et al.~2018) of the HSC SSP (Aihara et al.~2018a).\\
\end{tabnote}
\end{table*}

The CHORUS filter set consists of four NBs (NB387, NB527, NB718, and NB973) and one IB (IB945) in addition to NB921 from HSC SSP.
Table~\ref{tab:filters} is a summary of the characteristics of the filters and the targeted spectral features of high-$z$ galaxies and the redshift ranges.
Fig.~\ref{fig:filters} shows the total efficiency of the NBs as well as HSC BBs.\footnote{In the UD COSMOS field of the HSC SSP, there are two filters each for the $r$- and $i$-bands, namely $r2/r$ and $i2/i$ \citep{Kawanomoto2018}. 
The newer $r2$ and $i2$ filters were used about 60\% in the data according to the HSC PDR2 web page (https://hsc-release.mtk.nao.ac.jp/doc/index.php/data-2/). 
These two filter images were all combined to make final coadd images of the $r$- and $i$-bands \citep{Aihara2019}.}
The CHORUS filter set not only covers a wide redshift range of $2\lesssim z\lesssim 7$ of Ly$\alpha$ but also enables us to observe multiple spectral features at specific redshifts at the same time.
As found in Fig.~\ref{fig:filters}, the LyC and He~{\sc ii} 1640 line for $z\simeq3.3$ LAEs selected with NB527 can be observed with NB387 and NB718, respectively, and similarly, the LyC, C~{\sc iv} line, and He~{\sc ii} line for $z\simeq4.9$ LAEs selected with NB718 can be observed with NB527, NB921, and NB973, respectively.
For $z\simeq2.2$ LAEs selected with NB387, the He~{\sc ii} line can be observed with NB527 (see Table~\ref{tab:filters}).
The last CHORUS filter, IB945, is a filter to select LBGs within a redshift range similar to that of LAEs selected with NB921 ($z\simeq6.6$) and NB973 ($z\simeq7.0$) (see section~6.4 for its science case).
It is also useful to trace the line-free continuum between the C~{\sc iv} and He~{\sc ii} lines at $z\simeq4.9$ (see Fig.~\ref{fig:filters}).
In the following, we further describe the philosophy of the HSC SSP and CHORUS filter specifications and the characteristics of the CHORUS filters measured in the laboratory.

The NB filter set for the HSC imaging survey was designed very carefully \citep{Ouchi2018}.
The filter wavelengths in the redder wavelength range were determined to pass through OH sky windows at 816, 921, 973, and 1010 nm.
The three filters, except for the one at 973 nm, were used for the NB imaging observations in the HSC SSP survey \citep{Aihara2018a} called SILVERRUSH \citep{Ouchi2018}.
The remaining NB973 filter became a CHORUS filter.
The filter wavelengths in the blue wavelength range were determined to ensure that multiple spectral features of the LAEs selected by an NB filter were captured by the other NB filters and strong rest-frame optical emission lines of the LAEs passed through the OH sky windows in the near-infrared range maximally.
The selected wavelengths were 387, 527, and 718 nm, which became the CHORUS filters.\footnote{Although NB387 was also used for the `Deep' layer observations in the HSC SSP \citep{Aihara2018a,Ouchi2018}, there was no data taken in the footprint of the `UltraDeep' layer in the COSMOS field, where the CHORUS observations were performed.}
This combination allows us to observe LyC, Ly$\alpha$, and He~{\sc ii} at $2\lesssim z \lesssim 5$ as described above.
The [O~{\sc ii}] 3727 and H$\alpha$ lines for $z\simeq2.2$ LAEs selected with NB387 fall in the OH sky windows at 1.19 and 2.10 \micron.
The [O~{\sc iii}] 4959/5007 for $z\simeq3.3$ LAEs selected with NB527 are also observable at 2.15/2.17 \micron\ in the middle of the $K$-band.

The full-width-at-half-maximum (FWHM) of each filter was also determined carefully \citep{Ouchi2018}.
The FWHM of NB921 was first determined, to cover the full width of the OH sky window.
The FWHMs of the other NB filters, except for NB973 and IB945, were subsequently determined to ensure the same efficiency for the Ly$\alpha$ line detection as NB921.
Because the observed equivalent width of a line is $EW_{\rm obs}=(1+z)EW_0$, where $EW_0$ is the intrinsic equivalent width and $z$ is the source redshift, the FWHMs were determined by the relation, $FWHM=FWHM_{921}\times(1+z)/(1+6.58)$.
Note that 6.58 is the redshift of the LAEs selected with NB921.
NB973 has a narrower FWHM than the relation to enhance the efficiency of the Ly$\alpha$ line detection and compensate for the shallower depth in the band.
IB945 has an FWHM to fill the wavelength gap between NB921 and NB973.

The HSC filter is 600 mm in diameter, making it a challenging task to realize a uniform multi-layer interferometric coating to produce narrow bandpass filters with sharp cut-on and cut-off wavelengths.
Therefore, there was a small deviation from the specifications during the production process.
We examined the performance of the NB filters produced by the manufacturers, Optical Coatings Japan (NB387, NB527, NB718, NB921, and NB973) and Materion (IB945), using the method described in \citet{Kawanomoto2018}.
Slight non uniformities in the central wavelength (CW), FWHM, and transmission peak (TP) were found along the radial direction.
The CW tended to increase with the radius, except for NB387 where the CW variation was a complex way.
The FWHM also tended to increase in the outer part, except for NB387 where it decreased with the radius.
The overall CW variation was less than 0.3--0.5\% (the worst case was found in NB527), whereas the FWHM variation is ranging from a few--10\% (NB387).
The TP also tended to increase with the radius, except for NB387 again, where the TP decreased with the radius.
The overall TP variation was less than 1--10\% (NB973).
A higher degree of uniformity was found in the azimuthal direction.
Table~\ref{tab:filters} presents the CWs, FWHMs, and TPs for the area-weighted mean transmission functions of the CHORUS filters.
The values for the transmission functions at the center positions of the filters are also provided for reference.
The machine-readable filter transmission curves are available at the Subaru/HSC web site.\footnote{https://subarutelescope.org/Observing/Instruments/HSC/sensitivity.html}

\section{Observations, data reduction and photometric catalogs}\label{sec:obs}

\begin{table*}
  \tbl{Summary of CHORUS observations and data qualities.}{%
  \begin{tabular}{lllllll}
      \hline
      & & Exposure$^a$ & PSF FWHM$^b$ 
      & \multicolumn{2}{c}{$5\sigma$ depth$^c$ [AB]} & Area$^d$ \\
      Filter & Observation dates & [h] 
      & [$''$] & $\phi 1.5''$ & $\phi 2.0''$
      & [deg$^2$] \\
      \hline
      NB387 & Jan.\ 17, 18, 19, 2018 & 21/17.3 & 1.01 (0.99) & 25.80$^e$ (26.07)$^e$ & 25.42$^e$ (25.65)$^e$ & 1.561 \\
      NB527 & Dec.\ 16, 17, 18, 2017; Mar.\ 15, 16, 18, 2018 & 10.5/8.9 & 0.83 (0.82) & 26.72 (26.87) & 26.30 (26.46) & 1.613 \\
      NB718 & Feb.\ 25, Mar.\ 23, 25, 2017 & 13.5/7.7 & 0.69 (0.68) & 26.29 (26.47) & 25.87 (26.06) & 1.575 \\
      IB945 & Dec.\ 1, 2, 3, 12, 2018 & 10.2/10.2 & 0.62 (0.61) & 25.77 (25.92) & 25.36 (25.49) & 1.558 \\
      NB973 & Jan.\ 26, 28, 2017 & 15/14.7 & 0.64 (0.64) & 25.19 (25.37) & 24.79 (24.96) & 1.603 \\
      \hline
    \end{tabular}}\label{tab:obs}
\begin{tabnote}
$^a$ On-source exposure time/effective exposure time for the final coadd image, except for {\tt patch}es at the edge of the field-of-view, where these times are somewhat shorter due to dithering.\\
$^b$ Full-width-at-half-maximum (FWHM) of the point spread function (PSF): The area-weighted average value (the value in the central {\tt patch} 404).\\
$^c$ $5\sigma$ limiting magnitude for circular apertures of $1.5''$ or $2.0''$ in diameter: The area-weighted average value (the value in the central {\tt patch} 404).\\
$^d$ Effective area after removing masked regions.\\
$^e$ Value corrected for zero-point offset ($-0.45$ mag.). See an explanation in the last paragraph of section~\ref{sec:obs}.
\end{tabnote}
\end{table*}

We were awarded 13 nights in the classical mode of the Subaru Telescope for four consecutive semesters from S16B to S18A (i.e., 2 years) under a Subaru open-use intensive program, S16B-001I (PI: A.~K.~Inoue).
We were also awarded 11.5 hours in the queue mode in semester S18B under an open-use normal program, S18B-004 (PI: A.~K.~Inoue), to supplement the time loss due to the weather during the original four semesters.
Under the two programs, we performed deep imaging observations using our NB filters on the HSC on the dates listed in Table~\ref{tab:obs}.
The on-source exposure time as well as the effective exposure time used for the final coadd images are also listed in Table~\ref{tab:obs}.
Some of the observation dates were not under a photometric condition (i.e., low transparency), and we discarded a fraction of the observed data owing to bad quality.
The observing field was the COSMOS field \citep{Scoville2007}, which has the same footprint as the HSC SSP Ultra Deep (UD) layer \citep{Aihara2018a}.
Any sky region in the HSC SSP survey is defined by a set of two numbers of {\tt tract} and {\tt patch} \citep{Aihara2018b}.
Most of the HSC SSP UD COSMOS is covered by {\tt tract} number 9813.
There are $9\times9$ sub-regions --- {\tt patch}es --- in a {\tt tract}. 
Each {\tt patch} has a sky area of $12'\times12'$ and a {\tt tract} has that of $1.8^\circ\times1.8^\circ$.
The numbers of {\tt patch}es are expressed as 000 (south-east corner), 001, 002, ..., 008 (north-east corner), 100, 101, ..., 808 (north-west corner).
The {\tt patch} configuration in the UD COSMOS {\tt tract} 9813 is found in Fig.~\ref{fig:mask}.

The data reduction was conducted using the HSC SSP pipeline, {\tt hscPipe}, version 6.7, which is the same as that used for the SSP PDR2 data processing.
We used the obtained imaging data satisfying the following conditions: 
exposure time $=1200$ sec, seeing $<1.5''$, and transparency $>0.4$ for NB387, 
exposure time $=1200$ sec (but 3 frames of 468--1016 sec), seeing $<1.2''$, and transparency $>0.75$ for NB527, 
exposure time $=1200$ sec, seeing $<1.2''$, and transparency $>0.76$ for NB718, 
exposure time $=1200$ sec, seeing $<1.2''$, and transparency $>0.9$ for NB973, and 
exposure time $=720$ sec, seeing $<1.2''$, and transparency $>0.7$ for IB945.
{\tt hscPipe} generates several multiband photometric catalogs, which we call ``CHORUS PDR1 catalog.''
The source detection is based on the signal-to-noise ratio ($S/N$) within the PSF against the background noise fluctuation \citep{Bosch2018} and the detection criterion is $5\sigma$, the same as the SSP PDR papers \citep{Aihara2018b,Aihara2019}.
The readers should refer to pipeline paper \citep{Bosch2018} and the SSP PDR2 paper \citep{Aihara2019} for the detailed procedures of the data reduction, source detection, and photometry.
In the following, we describe some specific points for the CHORUS data release.

The photometric zero-points for the CHORUS NBs were determined based on the BB photometry of the Galactic stars in the Pan-STARRS1 (PS1) catalog \citep{Chambers2016} similar to the method for the HSC SSP \citep{Aihara2018b,Aihara2019}.
The color terms translating PS1 BB magnitudes to HSC NB magnitudes were estimated based on the Pickles stellar spectral templates \citep{Pickles1998}. 
This seems work well, except for NB387, where a systematic zero-point offset was found in \citet{Liang2020} (and see the last paragraph in this section).
The multiband photometry was conducted with the so-called {\it priority order} \citep{Bosch2018} of the filters, which was defined as $i,r,z,y,g$ for the BB filters, followed by the order of NB921, NB816, NB973, NB718, NB527, NB1010, NB387, and IB945 for the NB filters.
The priority order of the filters was used to determine the object positions.\footnote{This is the so-called {\it forced} catalog based on multiband photometry at the positions in the most prioritized band. 
The {\tt hscPipe} makes another catalog based on photometry at the positions in each band separately, called {\it meas} (or unforced) catalog \citep{Aihara2018b}. 
We do not deal with the {\it meas} catalog in this paper because the photometry may not give reliable colors owing to potentially different positions in different bands.}
For the line emitters detected in a single NB image but not detected in any other BB and NB images, photometry on the NB image was conducted at the positions of the detections in the NB image.
However, for the line emitters that were also detected in other BB (and NB) images, the photometry positions were forced to be the positions in the most prioritized band among the images in which the objects were detected.
In such cases, photometry on the NB image, which includes the line flux, can be underestimated if there are spatial offsets between the prioritized band and NB positions.
This would not be a problem for the bright objects that are detected well both in the prioritized band and NB images; however, it might be an issue for the objects that are marginally detected in the prioritized band and whose positions in the band have a large uncertainty.
Quantifying this effect would need a large set of pipeline simulations that was reserved for future line emitter analyses.

There are several different photometric magnitudes in the CHORUS PDR1 multiband catalog.
In this paper, we do not recommend using {\tt cmodel} magnitudes, which represent the total flux densities obtained by the brightness profile fitting \citep{Bosch2018} and were considered representative magnitudes in SSP PDR1 \citep{Aihara2018b}.
This is because there is an issue of erroneously bright {\tt cmodel} magnitudes for some objects in SSP PDR2 \citep{Aihara2019} and a follow-up study \citep{Hayashi2020}.
According to section~6.6.3 of the SSP PDR2 paper \citep{Aihara2019}, this problem is likely to be caused by a failure in the deblending procedure in crowded regions of objects.
This problem can be severe in the HSC SSP UD footprint in the COSMOS field, where the CHORUS observations were performed, because the very deep BB depth yields a very high density of the detected objects.

We instead recommend using the {\tt undeblended{\_}convolvedflux} magnitudes for estimates of total magnitudes.\footnote{There is another similar measurement called {\tt convolvedflux}, for which the source deblender algorithm in {\tt hscPipe} works. 
On the other hand, the source deblending process was turned off for {\tt undeblended{\_}convolvedflux}.
We have found that the two magnitudes of {\tt undeblended{\_}convolvedflux} and {\tt convolvedflux} are identical in most cases.} 
These measurements are based on the aperture magnitudes of the final coadd images smoothed with a Gaussian function (see section 3.6 of the SSP PDR1 paper, \cite{Aihara2018b}).
The Gaussian smoothing size is defined as the resultant FWHM of the point sources in the smoothed images, which is $0.59''$, $0.84''$, $1.1''$, or $1.3''$ \citep{Aihara2019}.
This process produces a set of the smoothed BB and NB images with the selected same PSF FWHM.
Subsequently, the flux densities of each object in all the smoothed (i.e., PSF-matched) images are measured in a circular aperture of diameter size $1.1''$, $1.5''$, $2.0''$, or Kron-size, and aperture corrections are also applied assuming a point-source.\footnote{The amount of the aperture correction can be found as {\tt undeblended{\_}convolvedflux{\_}X{\_}Y{\_}apcorr} in the catalog database, where X and Y are the numbers indicating the target convolution size and the aperture size, respectively.}
This process is performed for all the detected objects as a function of {\tt hscPipe} \citep{Aihara2018b,Aihara2019}.
These PSF-matched magnitudes yield better colors in crowded fields \citep{Aihara2018b}, 
which is the most important feature to reliably select LAEs in two-color diagrams.
This procedure is very similar to the method performed in previous studies with {\tt SExtractor} \citep{sextractor}.
In this study, we adopt $1.5''$ ($2.0''$)-diameter apertures on $0.84''$ ($1.1''$)-smoothed images.\footnote{If the native PSF of a band is larger than the target smoothing FWHM, any Gaussian smoothing is not applied to the band and simply the native images are used for the aperture photometry (see section 3.6 of \cite{Aihara2018b}). Such a case happened for NB387 in the $0.84''$ target FWHM smoothing.}

There are two significant known issues in the SSP PDR2 photometric catalog \citep{Aihara2019}, and they are prevalent in the CHORUS PDR1 catalog also produced by {\tt hscPipe} 6.7.
The first one is that the uncertainties of the PSF-matched magnitudes are underestimated (section 5.8.11 of the SSP PDR1 paper, \cite{Aihara2018b}).
This occurs because covariances are introduced by the Gaussian smoothing process.
This problem persists in the SSP PDR2 release \citep{Aihara2019}.
Therefore, we do not recommend using these cataloged values of uncertainties for the PSF-matched magnitudes.
Instead, we estimated the uncertainties by measuring the rms values of the aperture flux densities in the actual images.
The limiting magnitudes provided in the next section are based on these measurements.\footnote{The uncertainty in a magnitude is given by $\delta m=(2.5/\ln10)/(S/N)$. The signal-to-noise ratio, $S/N$, of each object can be estimated from the ratio of {\tt undeblended{\_}apertureflux} in the catalog database to the $1\sigma$ aperture flux density calculated from the limiting magnitudes in the next section.}
The second issue is that there is a systematic offset in the zero-point of NB387 of the SSP PDR2 release \citep{Aihara2019}.
This is due to the metallicity effect of the 4,000 \AA\ break in stellar spectra on $NB387-g$ color. 
{\tt hscPipe} uses the Pickles spectral templates based on Solar metallicity stars \citep{Pickles1998} for zero-point estimations. 
However, the actual stars used for the calibration seem to be dominated by metal-poor halo stars because these stars are faint ($g\sim20$) and the HSC survey fields are located in high Galactic latitudes.
This template mismatch causes a systematic offset of the zero-point only for NB387.
A complete description and full analyses are found in appendix of \citet{Liang2020}.
We did not correct the NB387 magnitudes in the CHORUS PDR1 catalog but recommend users to apply a 0.45 mag subtraction correction to the NB387 magnitudes in the catalog, as recommended by the HSC SSP team.\footnote{https://hsc-release.mtk.nao.ac.jp/doc/index.php/known-problems-2/{\#}hsc-link-10}

\section{Imaging data quality}

\subsection{Definition of masked areas}\label{sec:mask}

\begin{figure}
  \begin{center}
    \includegraphics[width=70mm]{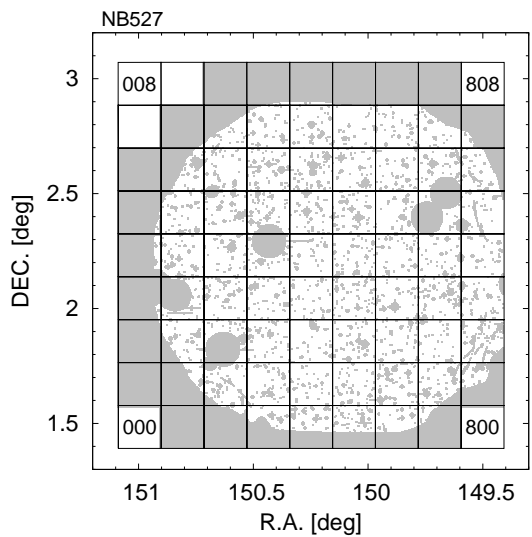}
  \end{center}
  \caption{Masked area (gray shade) in NB527 image as an example. HSC SSP UD COSMOS, {\tt tract} 9813, is divided into $9\times9$ {\tt patch}s as indicated by lattice. {\tt patch} numbers are indicated at four corners.} \label{fig:mask}
\end{figure}

We defined the masked areas in each CHORUS NB image based on the flags from {\tt hscPipe} and the visual inspection conducted by one of the authors.
Fig.~\ref{fig:mask} shows the masked area of the NB527 image as an example.
Pixels in each image that satisfied the following conditions of the {\tt hscPipe} flags were selected as masked pixels: 
{\tt pixelflags{\_}bright{\_}object} is {\tt True} (pixels affected by bright objects) or {\tt pixelflags{\_}saturatedcenter} is {\tt True} (pixels affected by count saturation).
In addition to these pixels, we also masked the pixels affected by the halos of bright stars out.
Generally, these halos are more prominent in NB images than in BB images probably owing to the multilayer interferometric thin film coating of NBs.
We supplementary set larger circular (sometimes elliptical) masks to cover these bright stars' halos if the pixels with the above flags do not cover sufficiently large areas.
One of the authors (S.Y.) selected these pixels by eyes.
The edge of the FoV was also defined by eyes by the same author to avoid lower S/N regions.
The effective survey area of each NB was calculated from the total number of non masked pixels in the image and is listed in Table~\ref{tab:obs}.
The exact positions of the masked pixels are available as image files in a FITS format.

\subsection{Sizes of point spread functions}

\begin{figure*}
  \begin{center}
    \includegraphics[width=80mm]{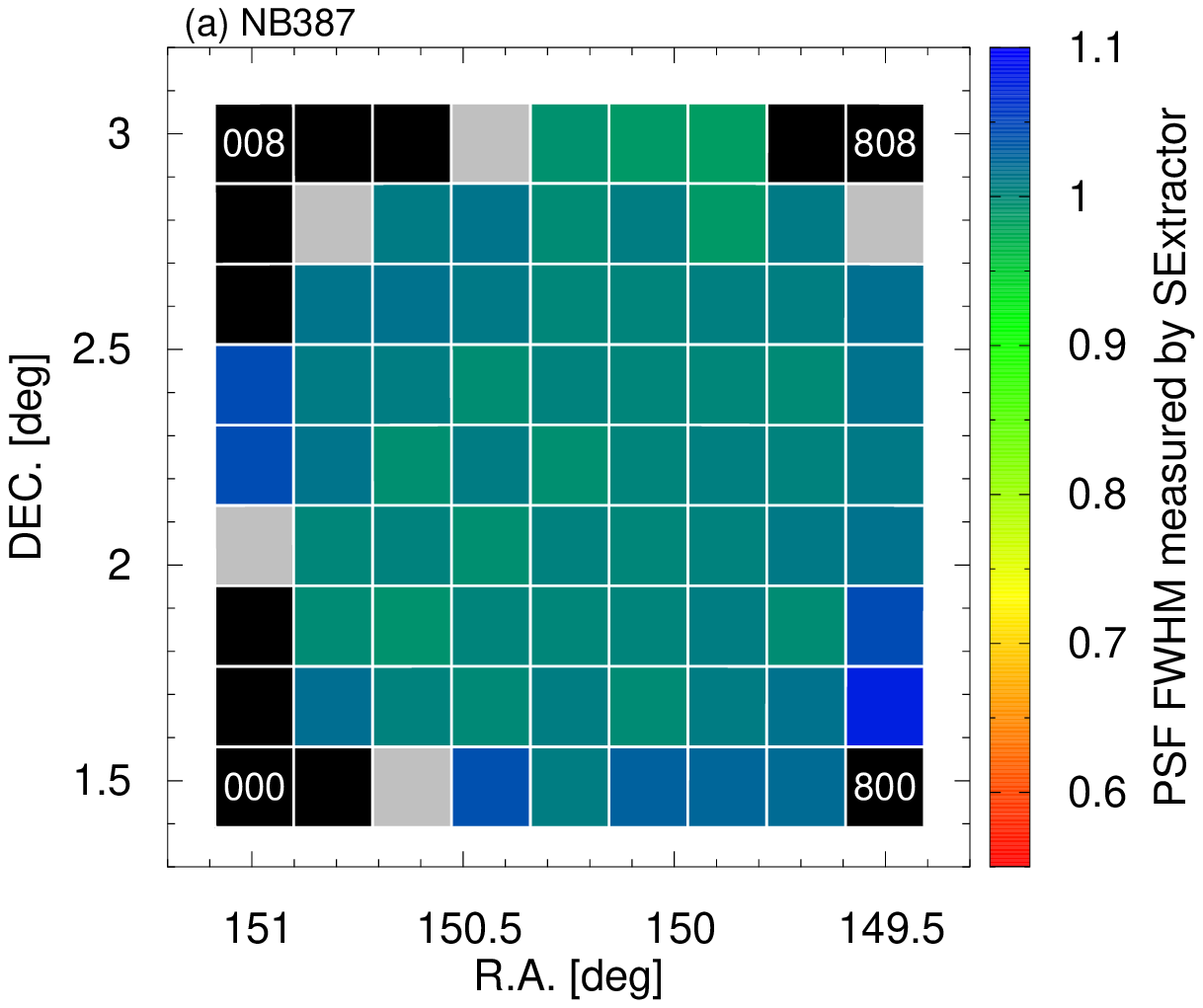}
    \includegraphics[width=80mm]{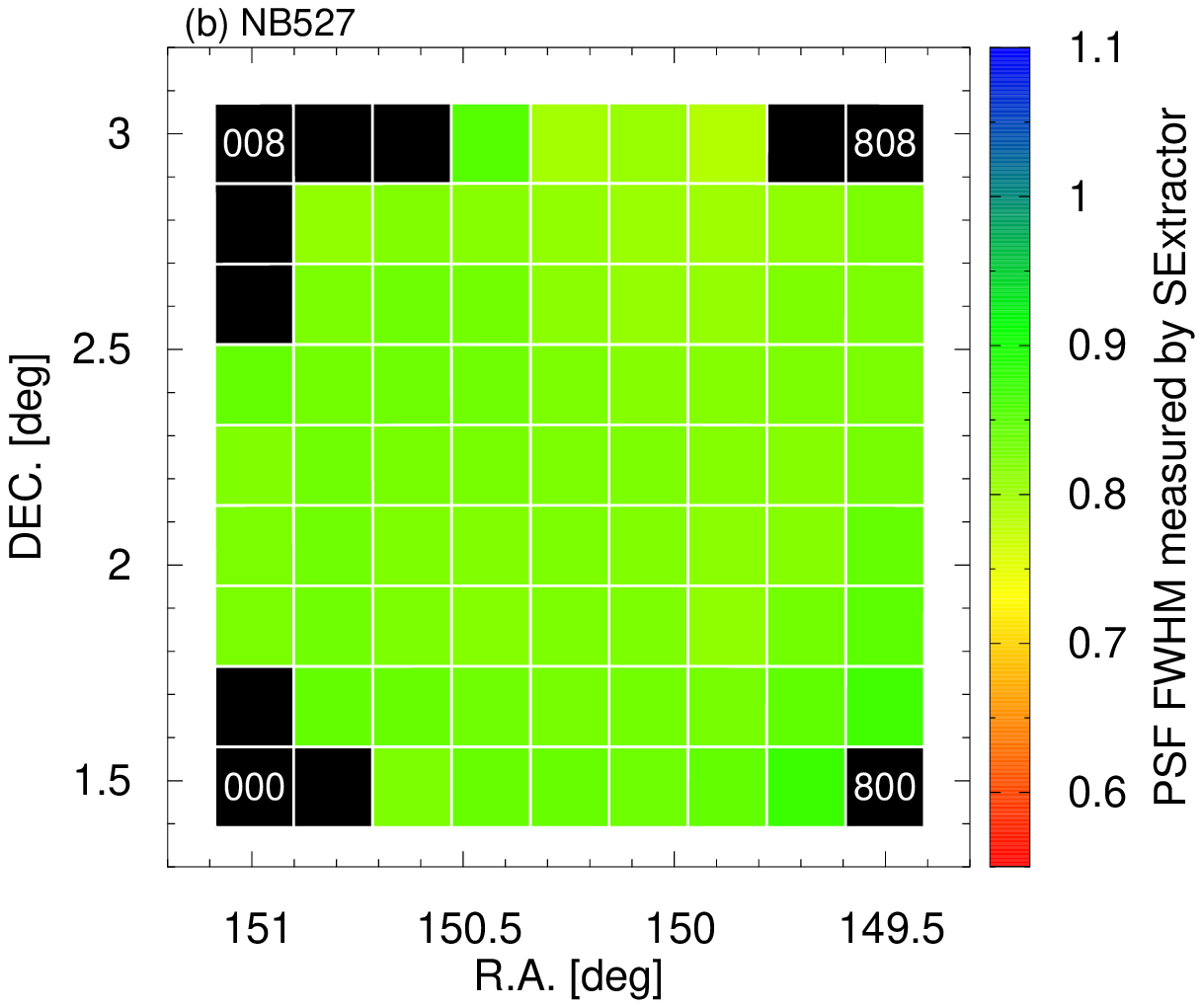}
    \includegraphics[width=80mm]{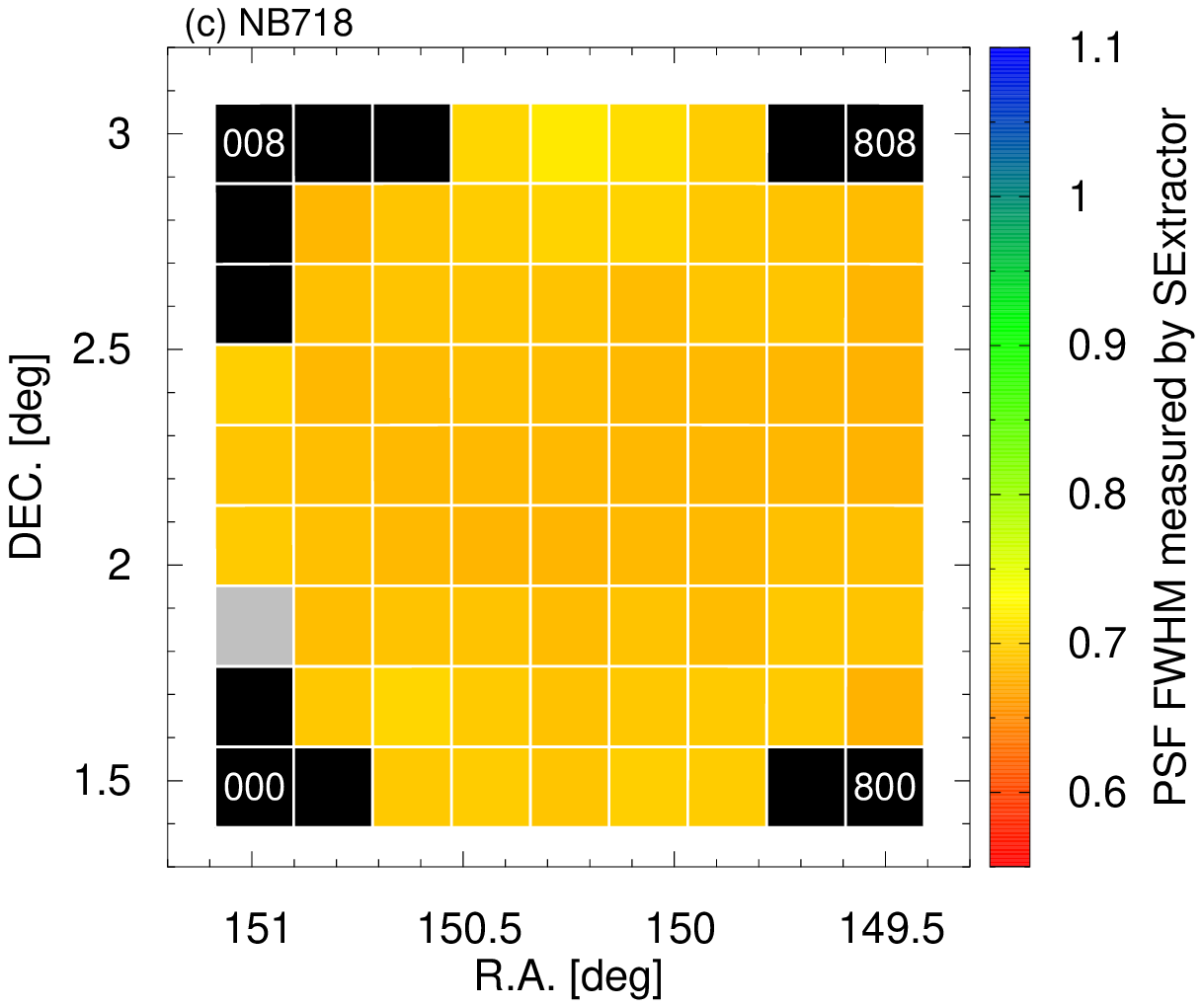}
    \includegraphics[width=80mm]{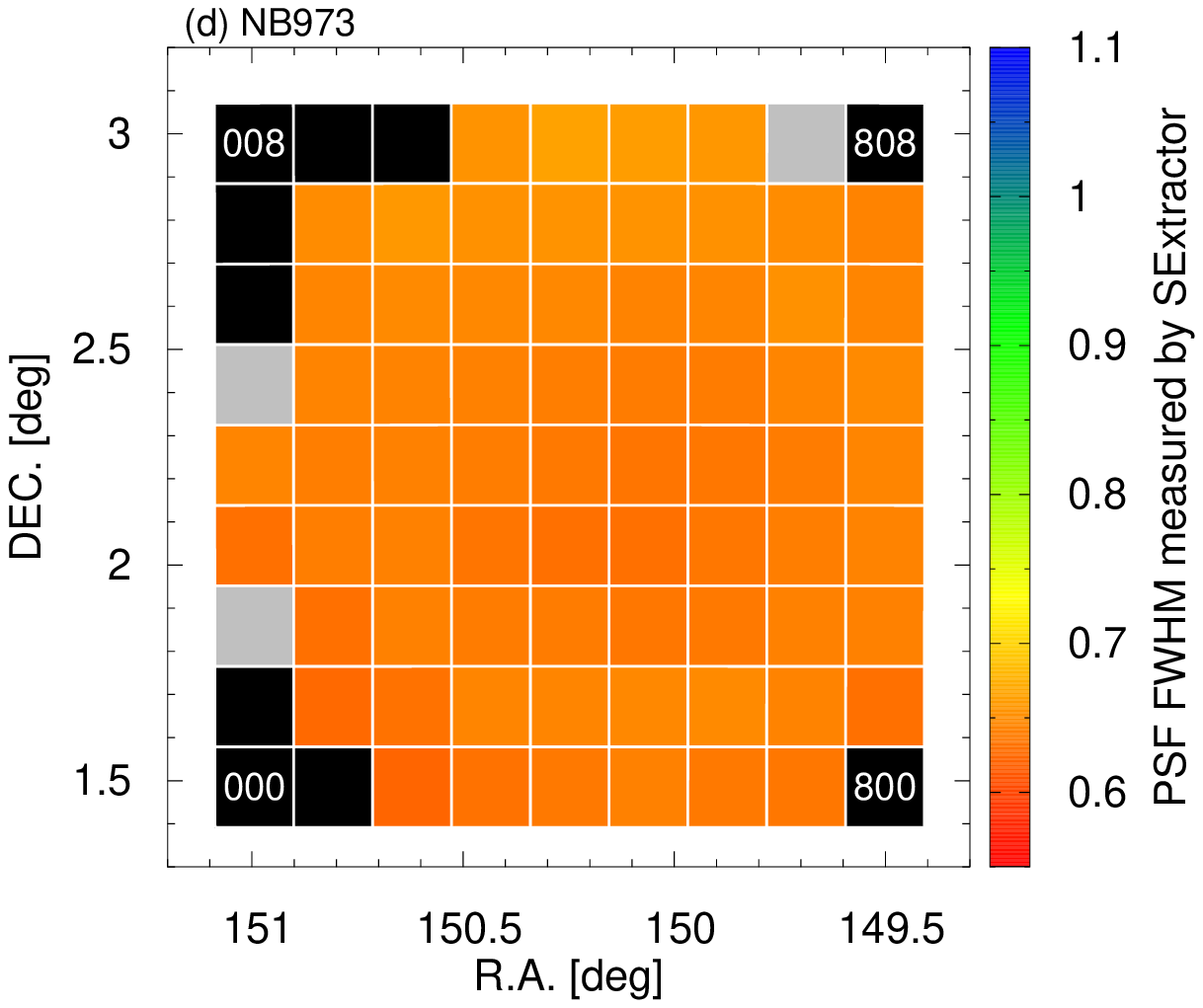}
    \includegraphics[width=80mm]{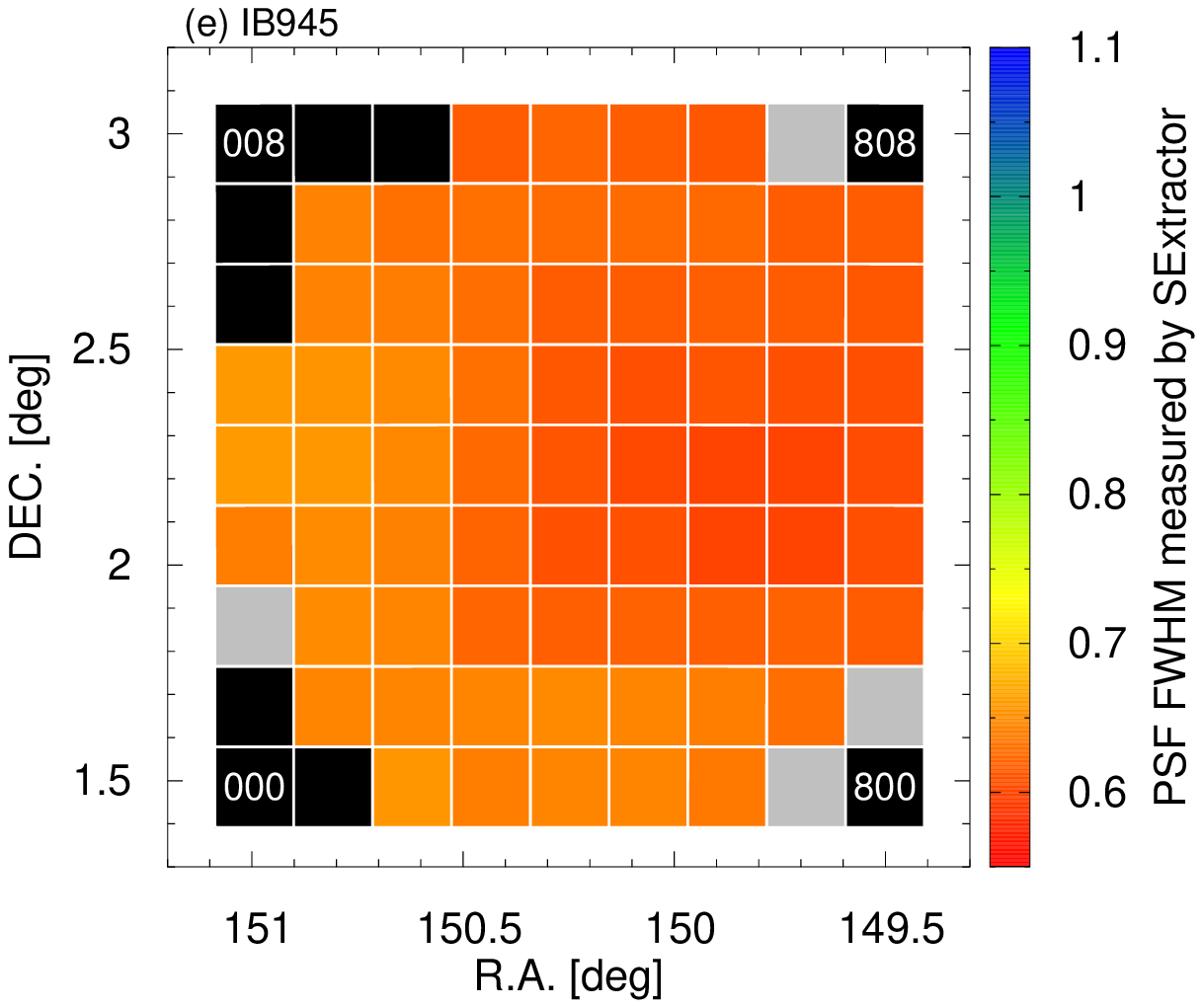}
  \end{center}
  \caption{FWHMs of PSF and their spatial variation for NB387 (top left), NB527 (top right), NB718 (middle left), NB973 (middle right), and IB945 (bottom). Black areas indicate {\tt patch}es that have no effective area. Grey areas indicate {\tt patch}es where there are insufficient number of stars for PSF FWHM measurement owing to small effective area.} \label{fig:psf}
\end{figure*}

We measured a representative FWHM of the PSF in each {\tt patch} of the CHORUS NB images using {\tt SExtractor}\footnote{http://www.astromatic.net/software/sextractor} ver 2.19.5 \citep{sextractor}.
We adopted a median of the {\tt FWHM{\_}IMAGE} measured by {\tt SExtractor} for stellar-like objects, which were not in the masked areas, as the representative FWHM of the PSF.
Stellar-like objects were selected depending on the measurements of {\tt SExtractor} as those satisfying the following three criteria: (1) {\tt MAG{\_}AUTO} in a range of [$18.0, 22.0$] for NB387 and NB527, [$18.0, 21.0$] for NB718 and NB973, and [$19.0, 21.0$] for IB945, (2) {\tt ELONGATION} $< 1.2$ (i.e., nearly circular shape), and (3) {\tt FLAG} $< 4$ (i.e., not located in a bad position, such as saturated or truncated pixels; see \cite{sextractor} for the details). 
If there were insufficient number of stellar-like objects in a {\tt patch} because of an extremely small non masked area, we did not measure the FWHM of the PSF.
The measured FWHMs and their spatial variations are shown in Fig.~\ref{fig:psf}, and the values of the area-weighted average and the central {\tt patch}, 404, are listed in Table~\ref{tab:obs}.
The FWHM values range from $1''$ to $0.6''$, depending on the band.
The spatial variation in the FWHM values in each NB is as small as $<0.02''$, demonstrating the excellent stability of the image quality across the entire FoV of the HSC.

\subsection{Limiting magnitudes}\label{sec:limitmag}

\begin{figure*}
  \begin{center}
    \includegraphics[width=80mm]{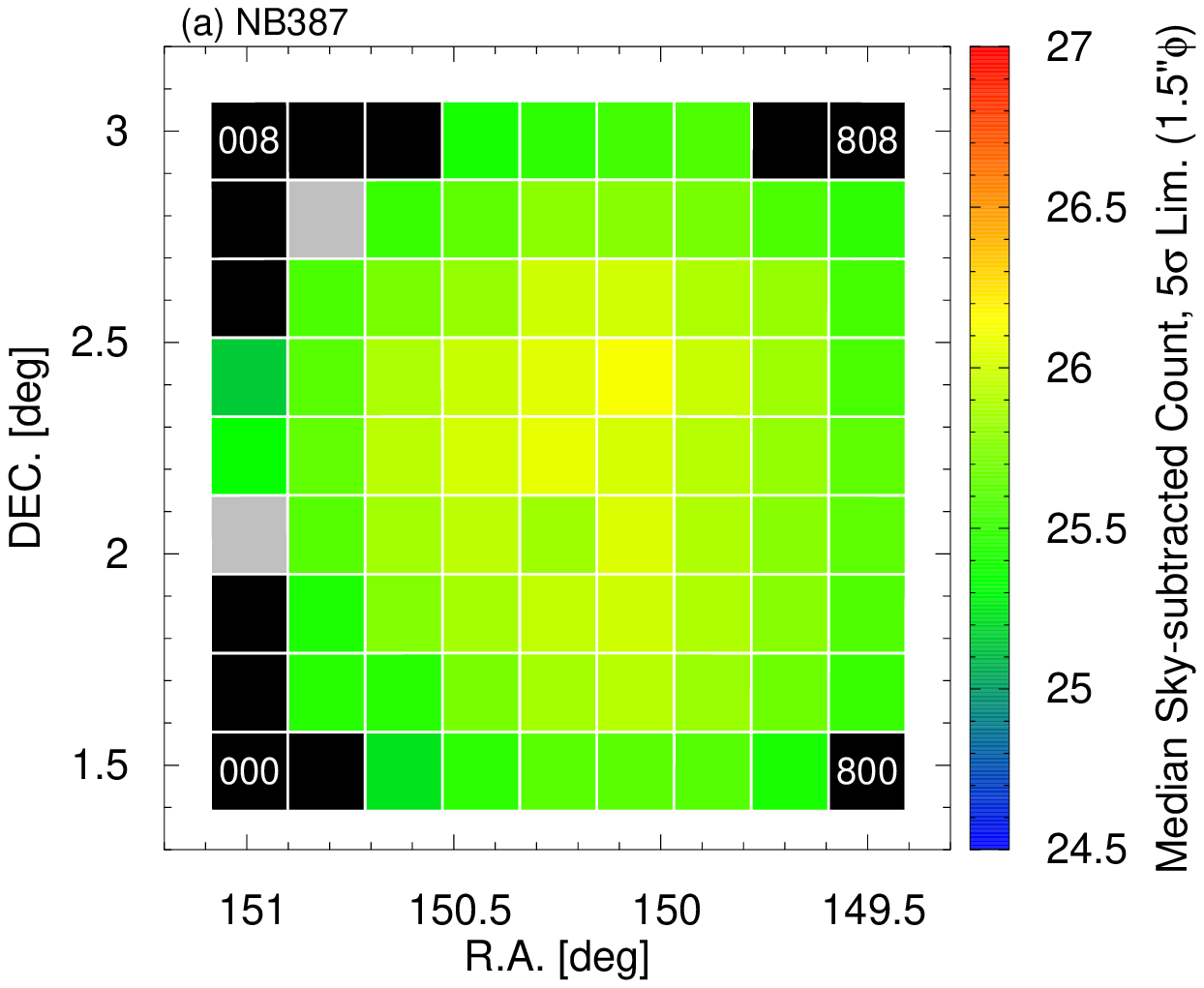}
    \includegraphics[width=80mm]{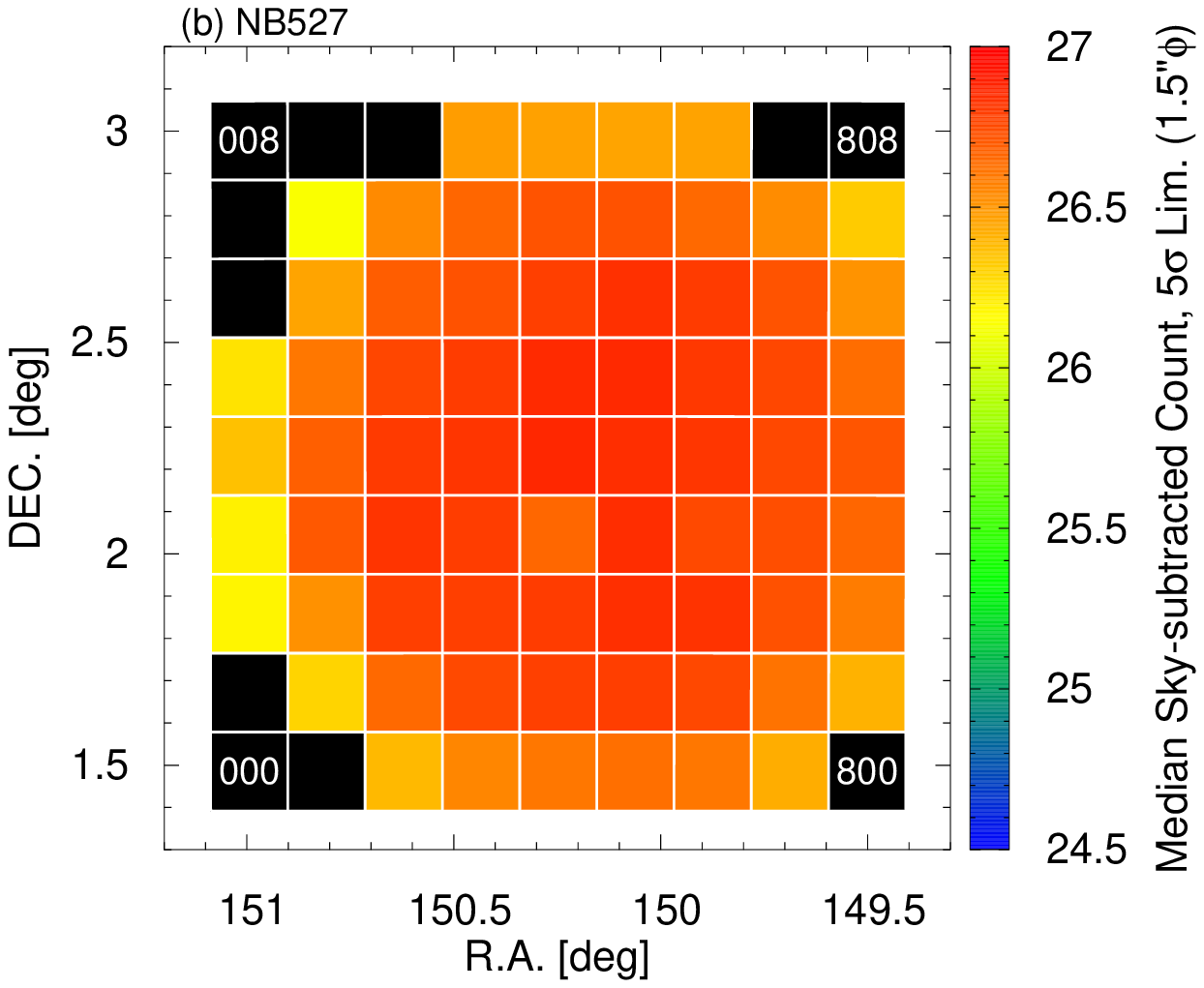}
    \includegraphics[width=80mm]{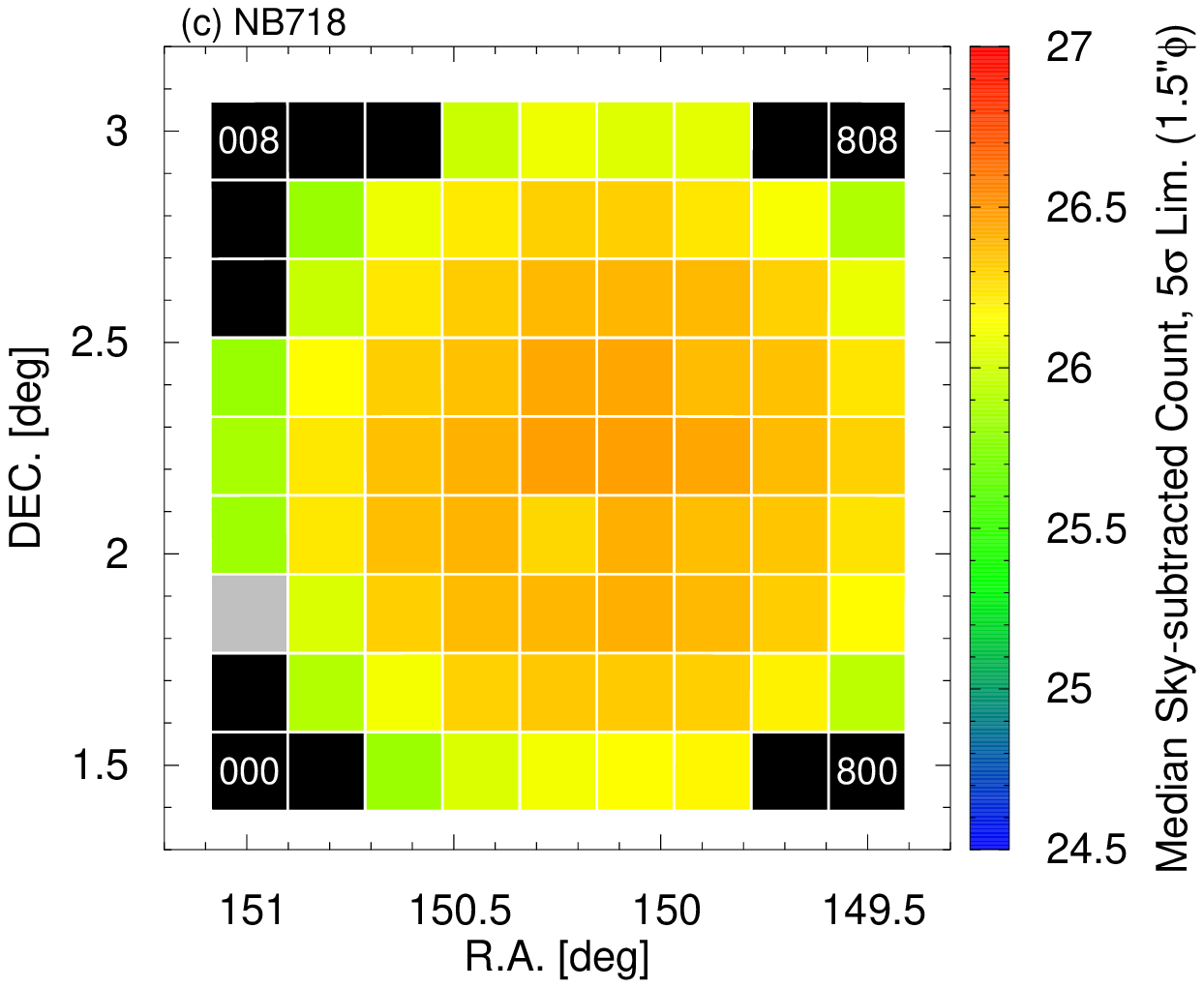}
    \includegraphics[width=80mm]{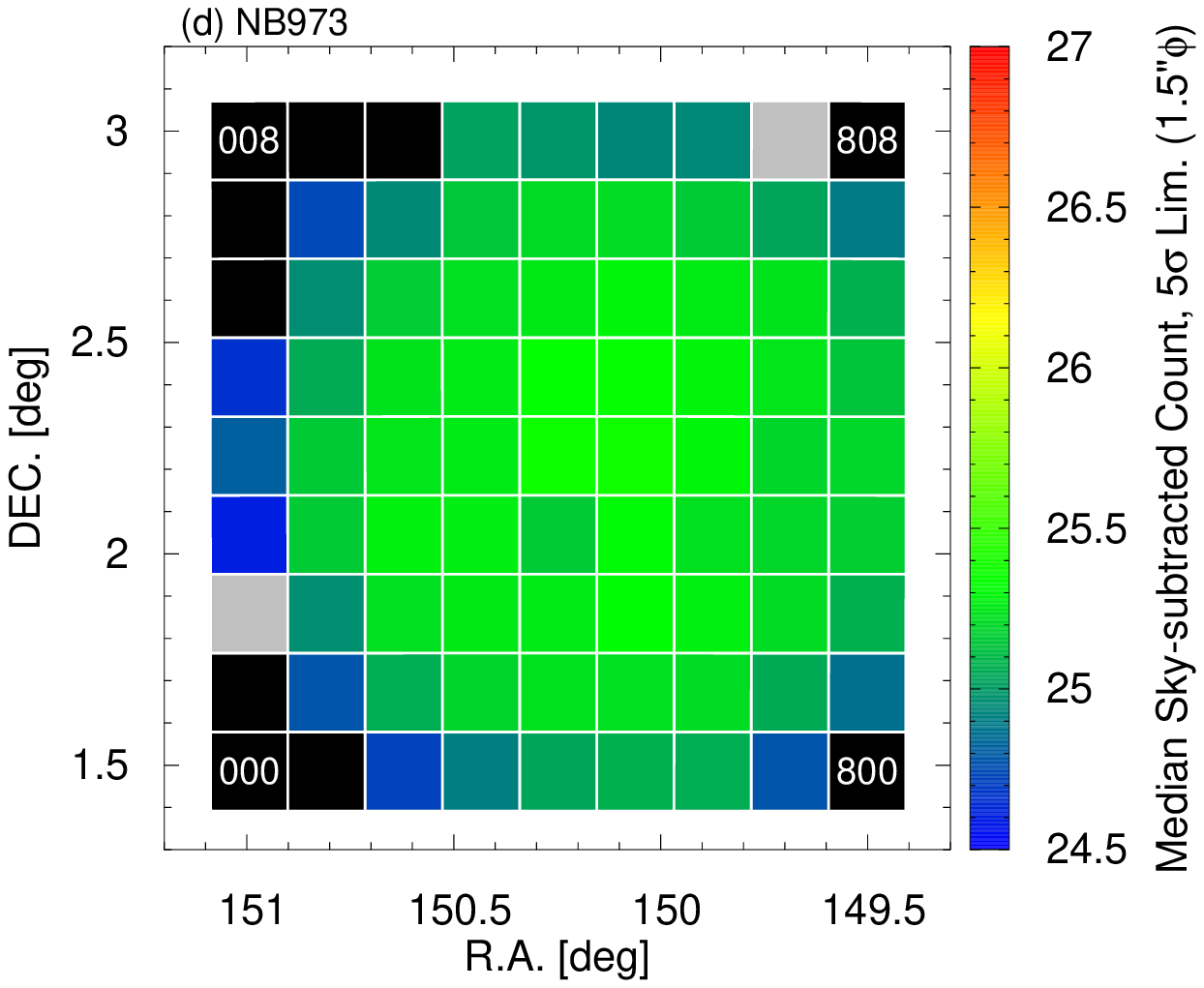}
    \includegraphics[width=80mm]{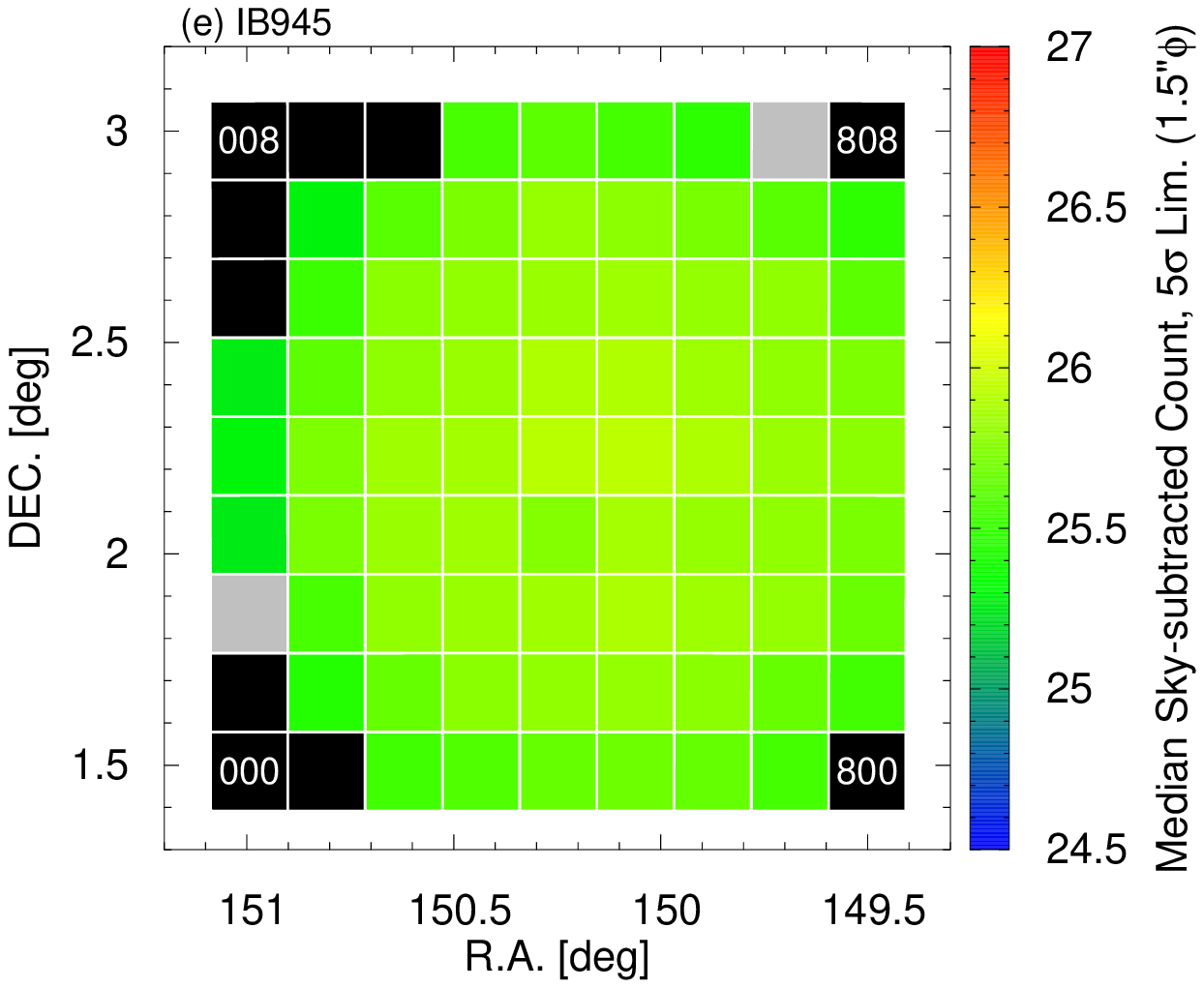}
  \end{center}
  \caption{Five-$\sigma$ limiting magnitudes and their spatial variation for NB387 (top left), NB527 (top right), NB718 (middle left), NB973 (middle right), and IB945 (bottom). Black areas indicate {\tt patch}es that have no effective area. Grey areas indicate {\tt patch}es where there are insufficient non-masked pixels for limiting magnitude measurements. Aperture size is $1.5''$ in diameter.} \label{fig:limitmag}
\end{figure*}

We measured the limiting magnitudes in each {\tt patch} of the CHORUS NB images by conducting aperture photometry at random sky positions with Python packages, 
{\tt Astropy}\footnote{https://www.astropy.org/}/{\tt photutils}\footnote{https://photutils.readthedocs.io/en/stable/}. 
A large number of random positions (10,000 as default, and 5,000 or 1,000 at the edge of the FoV depending on the available area) were prepared in each {\tt patch} by avoiding the masked pixels and the pixels of the objects detected by {\tt SExtractor} in advance. 
If the non masked area in a {\tt patch} was less than $1~\mathrm{arcmin}^2$, we did not measure the limiting magnitudes.
For convenience, we measured the limiting magnitudes with two aperture sizes ($1.5''$ and $2.0''$ in diameter) in each {\tt patch} of each CHORUS NB image.
We estimated the background brightness locally by adopting a median count in an annulus of $1.5''$ ($2.0''$) width and $2.5''$ ($3.0''$) inner diameter set around the $1.5''$ ($2.0''$) aperture and subtracted it from the aperture count as a correction for the background contribution.
The standard deviation, $\sigma$, was obtained from the histogram of the background-subtracted aperture counts by fitting a Gaussian function.
Note that we did not apply any aperture correction to the counts.
The $5\sigma$ values of the area-weighted average and those in the central {\tt patch}, 404, for the two aperture sizes are listed in Table~\ref{tab:obs}.
The spatial variation in the $5\sigma$ values for the $1.5''$-diameter aperture case are shown in Fig.~\ref{fig:limitmag}.
We can see a radial dependence of the limiting magnitudes; 
it is $\sim1$ mag shallower at the outer edge compared with that at the center.
The {\tt patch}, 403, shows a shallower depth than the radial trend, which was also reported in \citet{Hayashi2020}.
This is probably because a half of the four channels in a CCD chip (SDO-ID=0{\_}20/DET-ID=33) close to the {\tt patch} center was unavailable since November-December 2016 (see the CCD information page of HSC on Subaru Telescope website\footnote{https://www.naoj.org/Observing/Instruments/HSC/ccd.html}) and the dithering amount was too small to compensate the low-sensitivity area.

\subsection{Completeness}

\begin{figure*}
  \begin{center}
    \includegraphics[width=70mm]{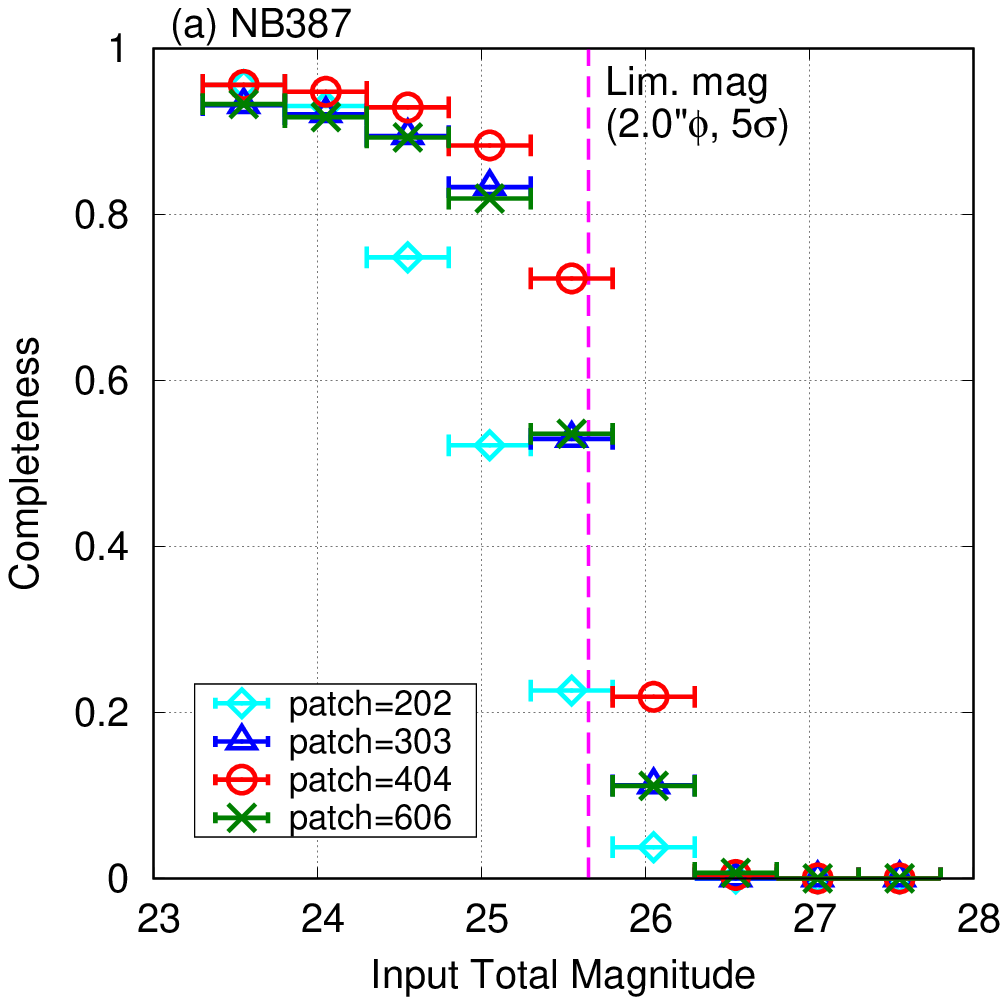}
    \includegraphics[width=70mm]{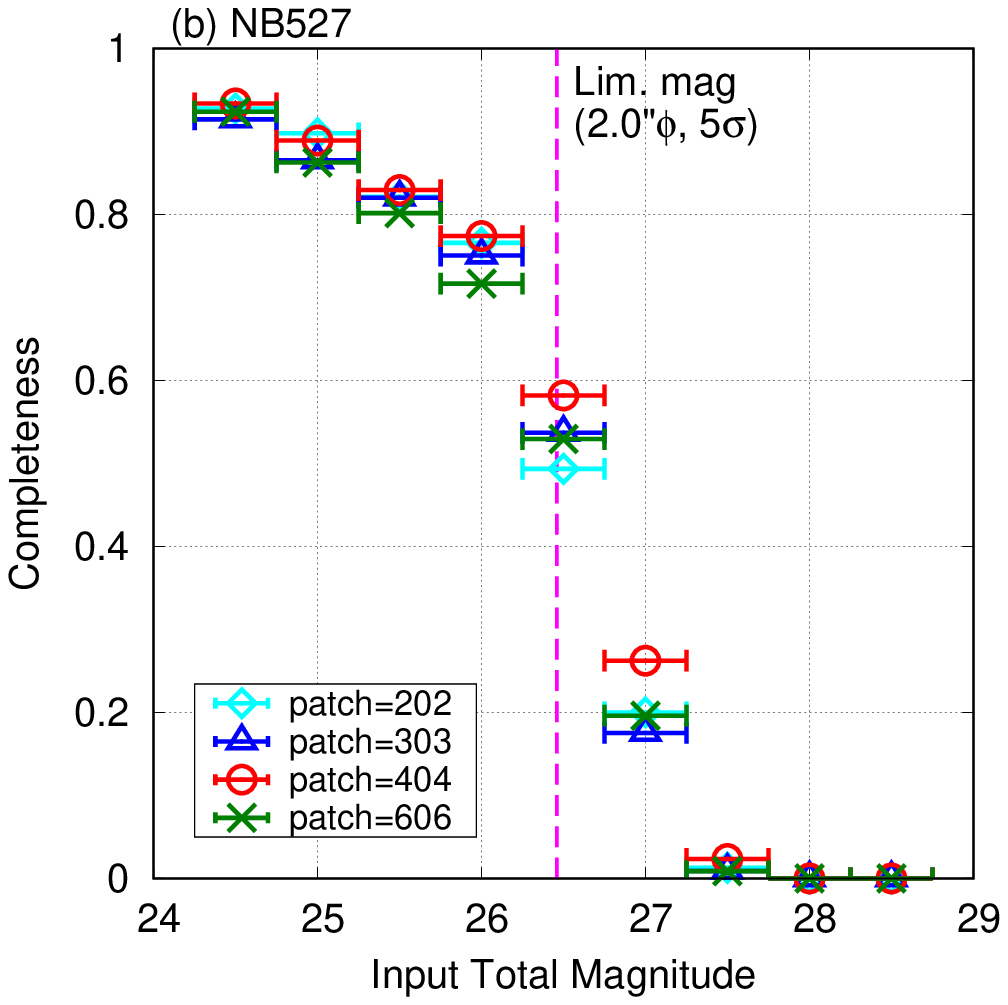}
    \includegraphics[width=70mm]{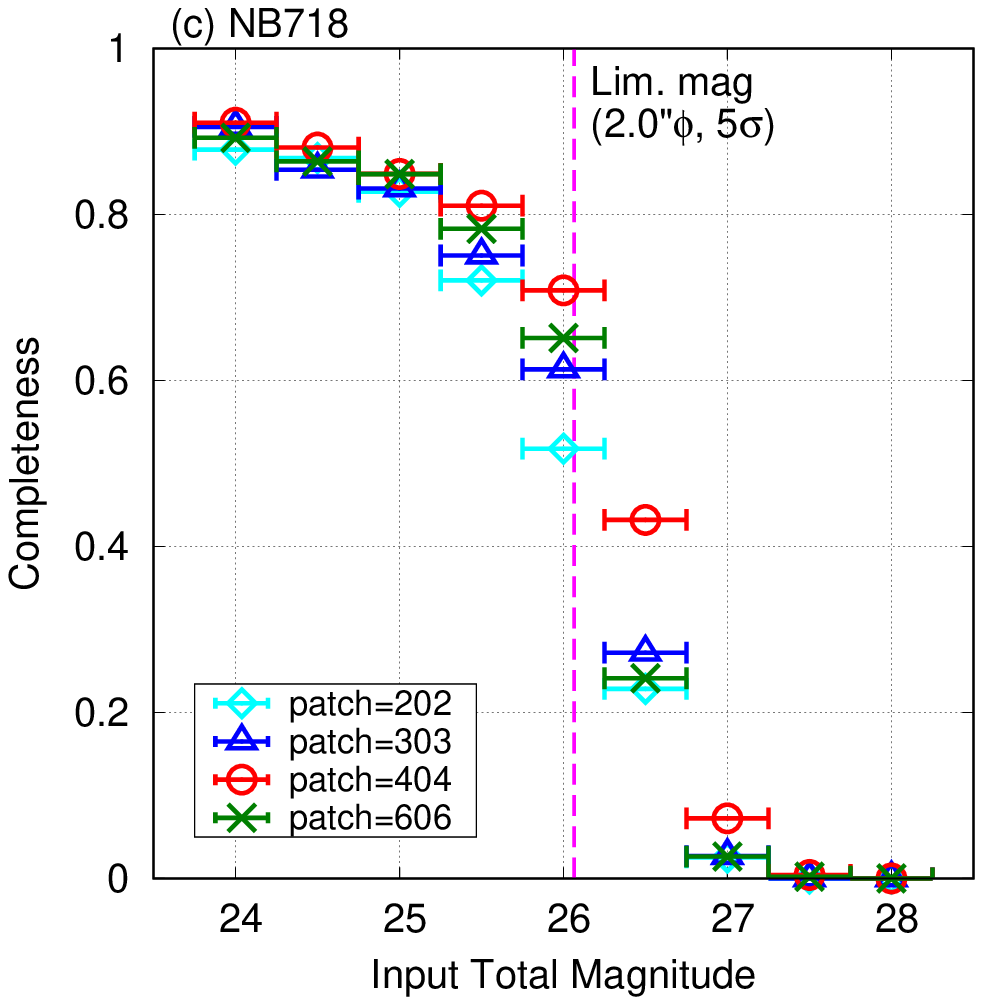}
    \includegraphics[width=70mm]{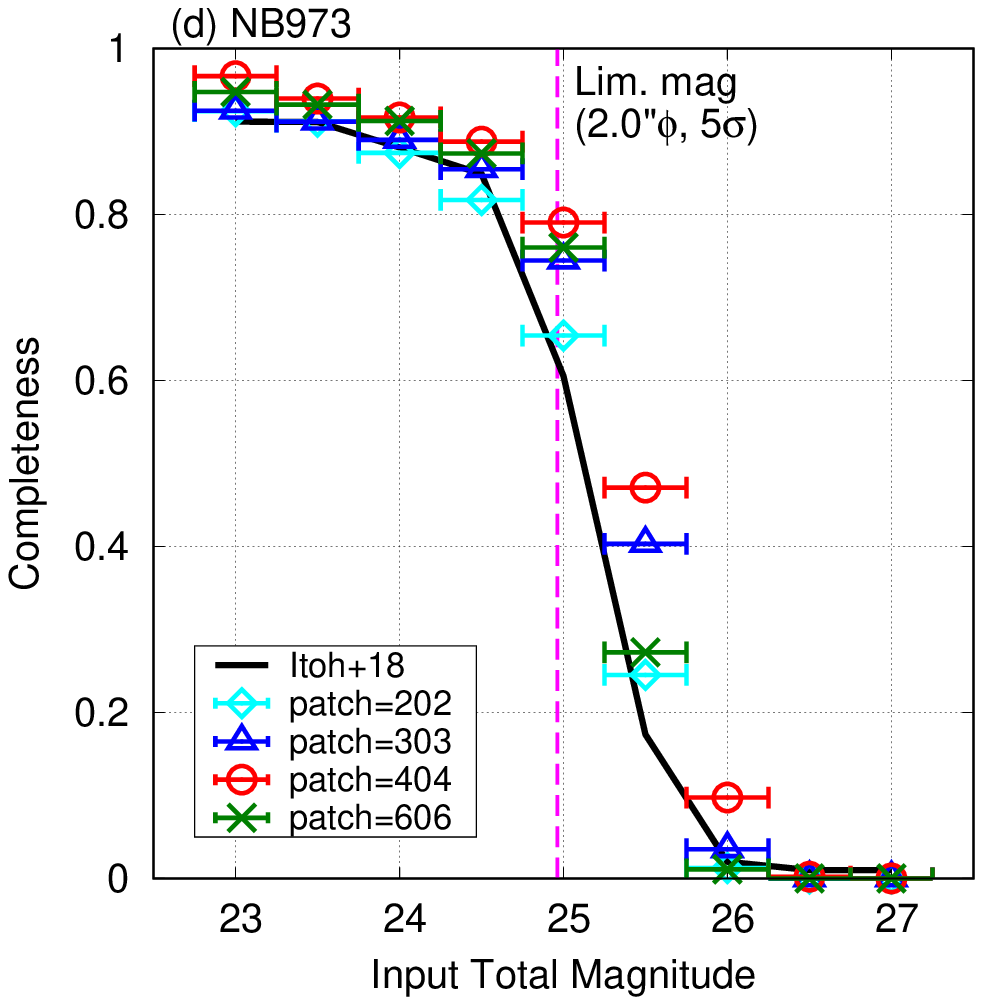}
    \includegraphics[width=70mm]{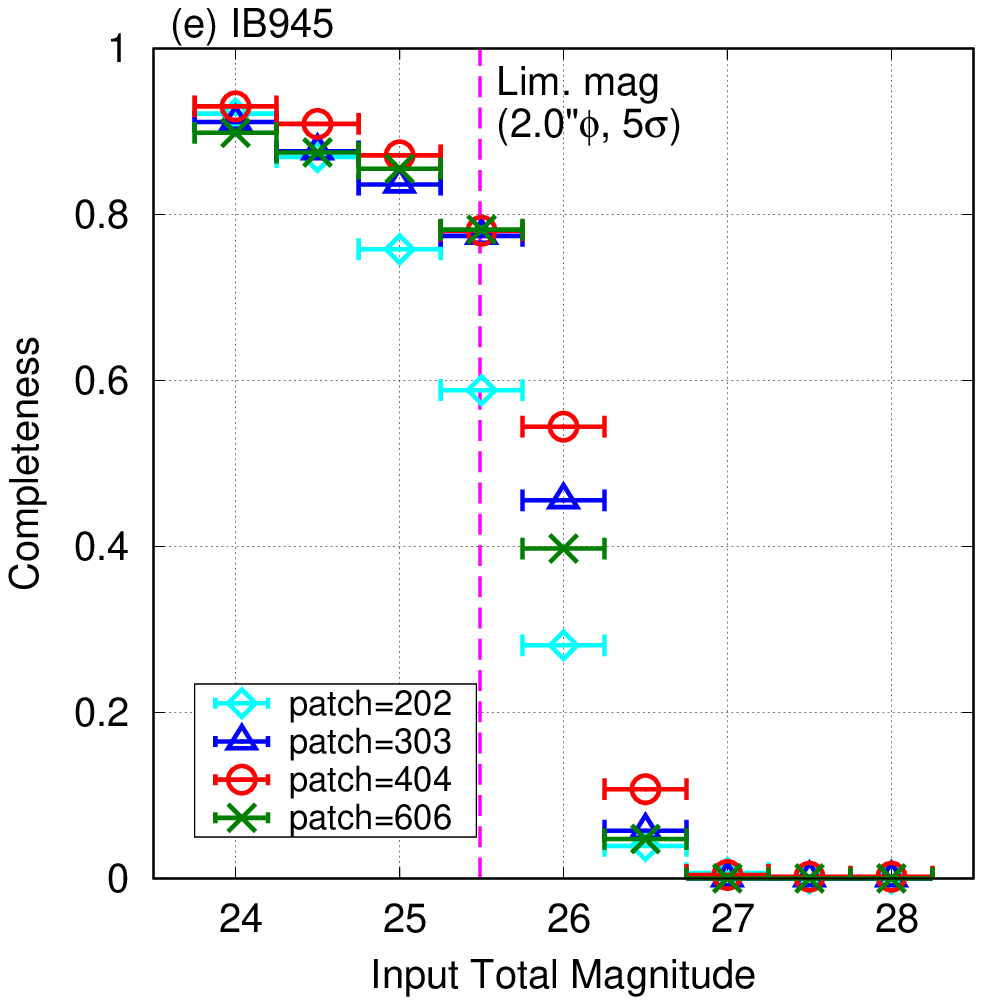}
  \end{center}
  \caption{Detection completeness of NB387 (top left), NB527 (top right), NB718 (middle left), NB973 (middle right), and IB945 (bottom). Cyan, blue, red, and green points with error-bars denote results for {\tt patch}es 202, 303, 404, and 606, respectively. In panel of NB973, black solid line indicates result of Itoh et al.~(2018). We also display $5 \sigma$ limiting magnitudes measured with $2.0''$-diameter aperture (not total magnitudes) in {\tt patch} 404 by vertical magenta dashed line in each panel. These results are obtained with {\tt undeblended{\_}convolvedflux} magnitudes; however, essentially same results are obtained with {\tt convolvedflux} magnitudes.} \label{fig:completeness}
\end{figure*}

We estimated the detection completeness based on a simulation of embedding and recovering numerous mock galaxies in actual CHORUS NB images using {\tt hscPipe}, Astropy, and {\tt GALSIM} \citep{GALSIM}\footnote{https://github.com/GalSim-developers/GalSim}.
These mock galaxies were assumed to have an intrinsic brightness profile described by a circular Sersic profile with index $n=1.5$ and a half-light radius of $1$ pix of the HSC images ($0.17''$).
The adopted radius corresponds to $1.4$--$0.9$ physical kpc for $2<z<7$ and very consistent with those of typical LBGs/LAEs \citep{Shibuya2015,Kawamata2018}.
When embedding the mock galaxies into each {\tt patch} image, we convoluted the brightness profile with the PSF of the image.
Because the intrinsic sizes of the mock galaxies were very small compared to the PSF size, the embedded profiles were similar to the PSF but slightly extended.
The locations to embed the mock galaxies were chosen randomly, not avoiding the actual objects, to consider the cases in which the extended and bright galaxies overlap and hide faint ones.
We considered nine intrinsic magnitudes of the mock galaxies in a range of $\pm2.0$ magnitudes with a 0.5 magnitude step around the $5\sigma$ limiting magnitude.
The number of the input mock galaxies was 500 for each intrinsic magnitude in each {\tt patch} of each band.
We ran {\tt hscPipe} to detect the mock galaxies and measure the magnitudes.
The criteria for successful recovery were detection within a radius of $0.5''$ from the embedded location and a magnitude difference of less than 0.5 magnitude.
We adopted the {\tt undeblended{\_}convolvedflux} magnitude ($0.84''$ FWHM convolution and $\phi 1.5''$ aperture case, except for NB387, where $1.1''$ FWHM convolution and $\phi 2.0''$ aperture case is adopted) as the output total magnitude for Fig.~\ref{fig:completeness}; however, the results with the {\tt convolvedflux} magnitude were the essentially same.
Note that the {\tt undeblended{\_}convolvedflux} and {\tt convolvedflux} magnitudes are aperture-corrected by a PSF model and equivalent to the total magnitude for the point sources.
The detection completeness is defined as the ratio of the number of successfully recovered galaxies to that of embedded mock galaxies.
The mock galaxies placed in the masked areas were excluded from the analysis in this final step.
To moderate the computation time, we conducted this experiment in only 4 {\tt patch}es of 202, 303, 404, and 606, instead of all the 81 {\tt patch}es.

Fig.~\ref{fig:completeness} shows the results of the simulations of the detection completeness.
We find that the completeness is $\gtrsim 60 \%$ at the $5\sigma$ limiting magnitude ($\phi2.0''$ not aperture-corrected) in each {\tt patch}, which is reasonable.
Even at magnitudes brighter than the limiting magnitudes, the completeness is only 80--95\% and does not reach 100\% because of the hiding effect by bright objects.
For NB527, the completeness is relatively lower than that of other bands.
This seems to be caused by the source confusion due to the very deep depth with a moderate PSF size.
Indeed, the surface number density of the detected objects in NB527 up to a $5\sigma$ limit of 26.46 AB ($\phi 2.0''$) corresponds to an areal fraction of $\sim 13$\% even if we assume a circular area with a radius equal to the PSF FWHM size ($r = 0.82''$) per object.
For NB973, we also present the average completeness over all the {\tt patch}es estimated similarly but independently by \citet{Itoh2018}; 
it is slightly lower than ours but is broadly consistent.
Actually, \citet{Itoh2018} used {\tt hscPipe} version~4.0.5, and their $5\sigma$-limiting magnitude in the $1.5''$-diameter aperture of 25.0 was 0.4 magnitude shallower than that of this study.

\section{Number counts}

\begin{figure*}
  \begin{center}
    \includegraphics[width=56mm]{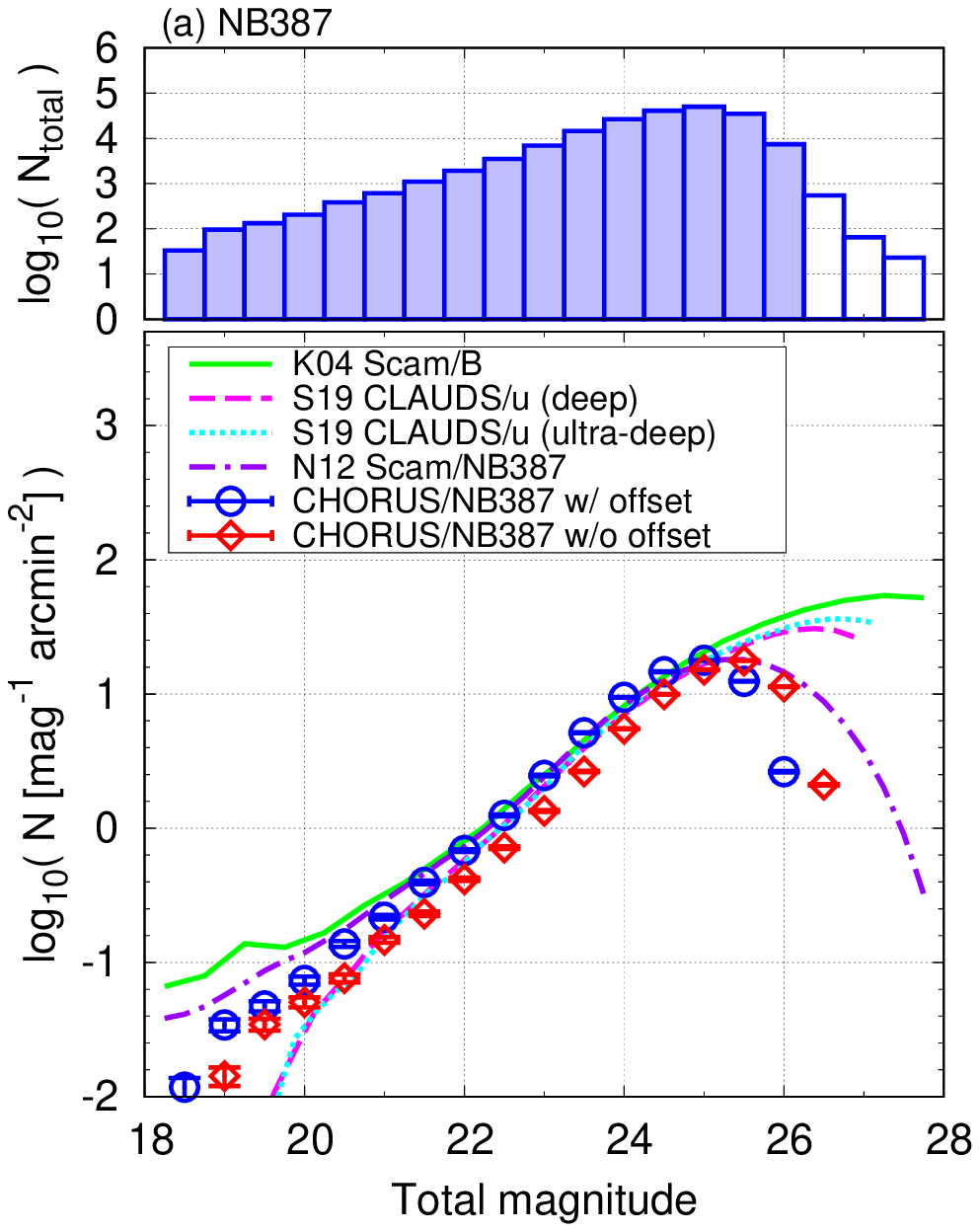}
    \includegraphics[width=56mm]{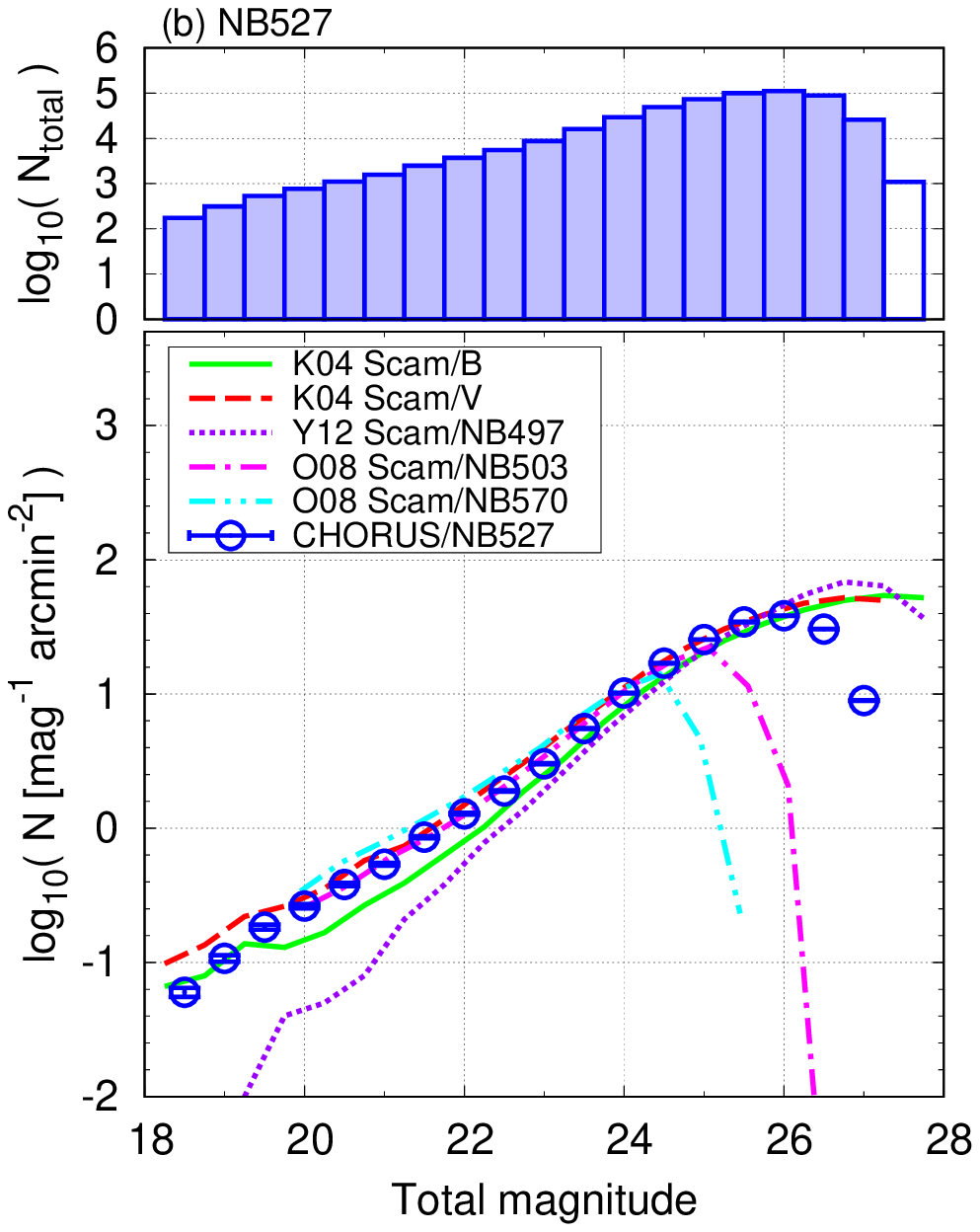}
    \includegraphics[width=56mm]{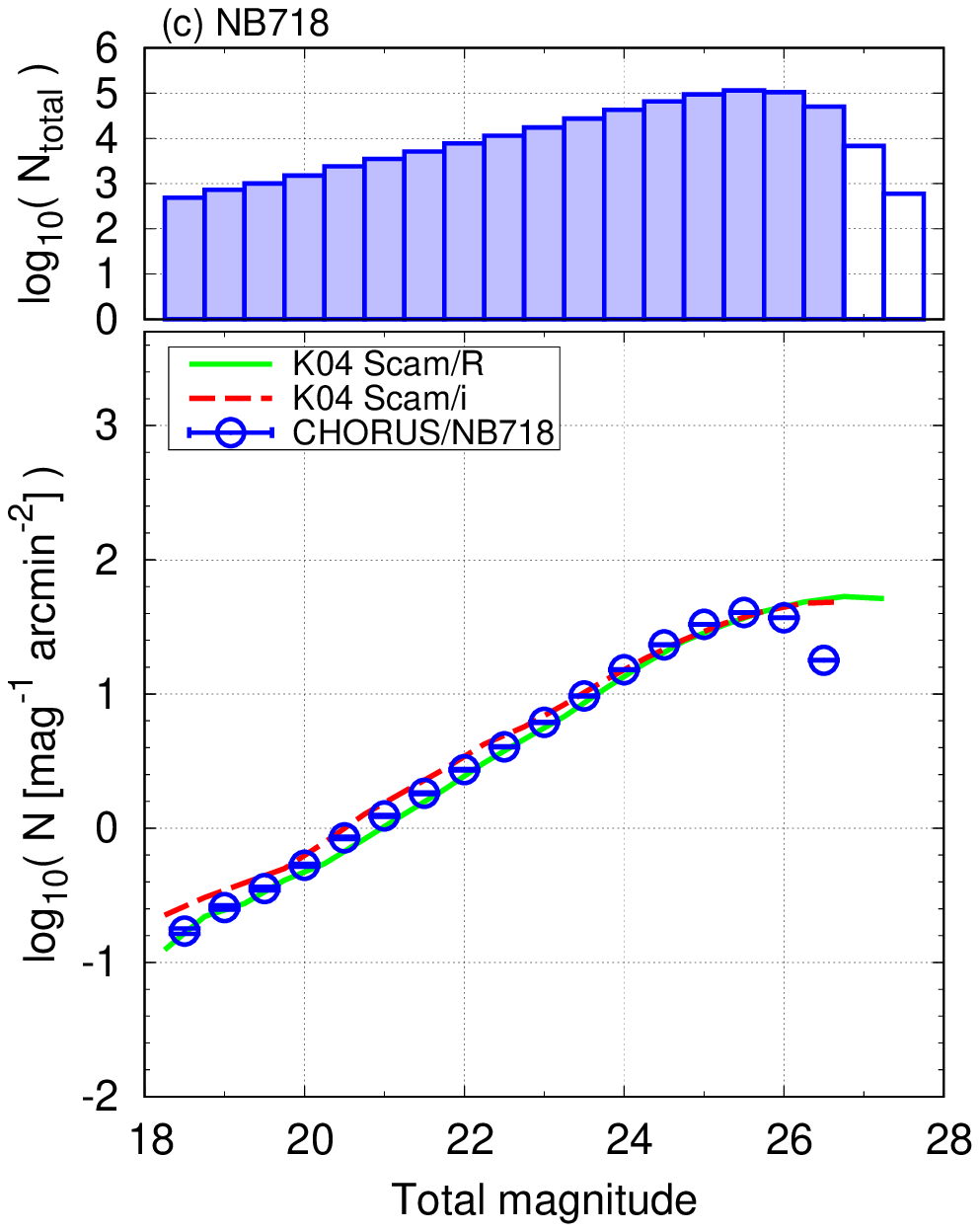}
    \includegraphics[width=56mm]{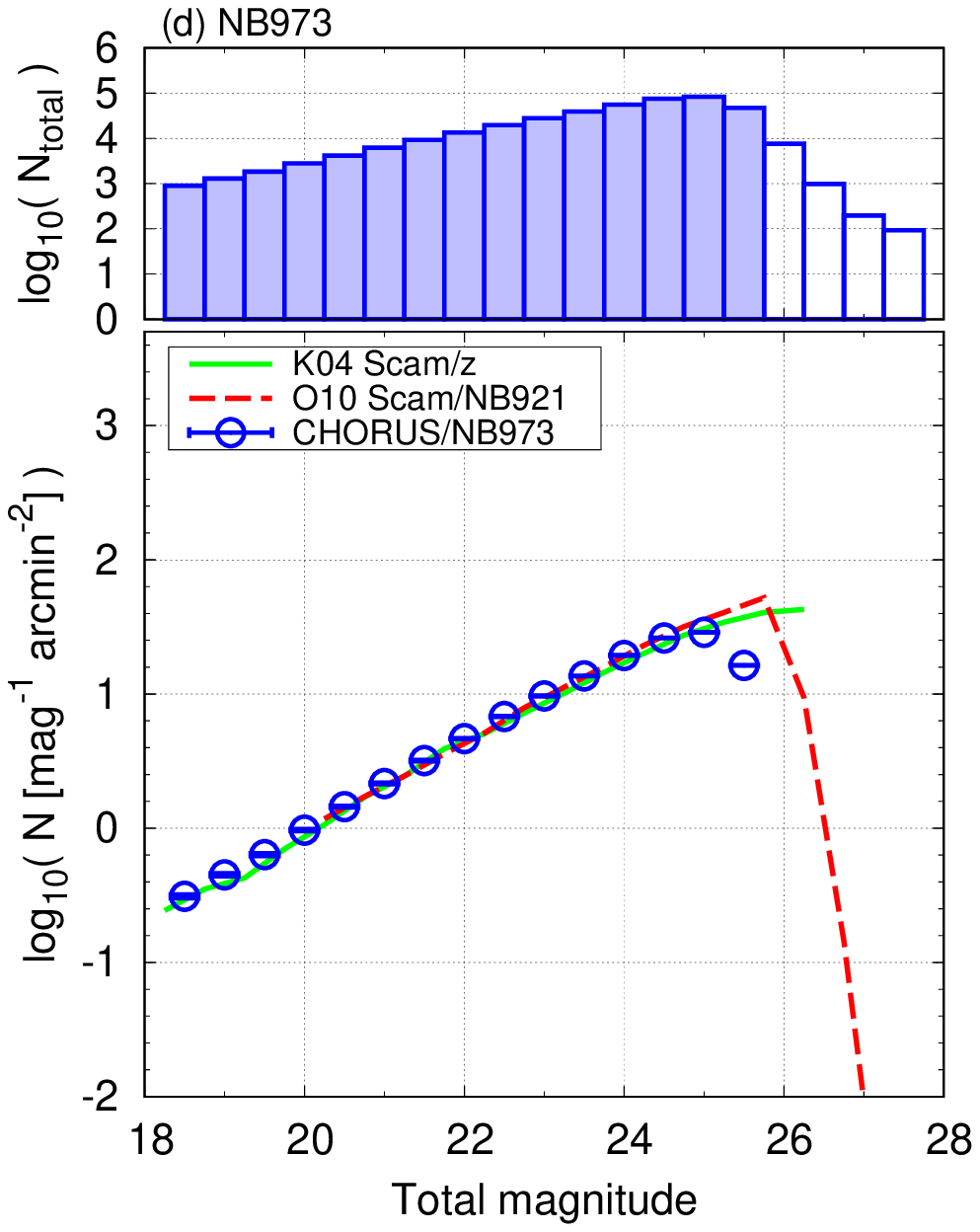}
    \includegraphics[width=56mm]{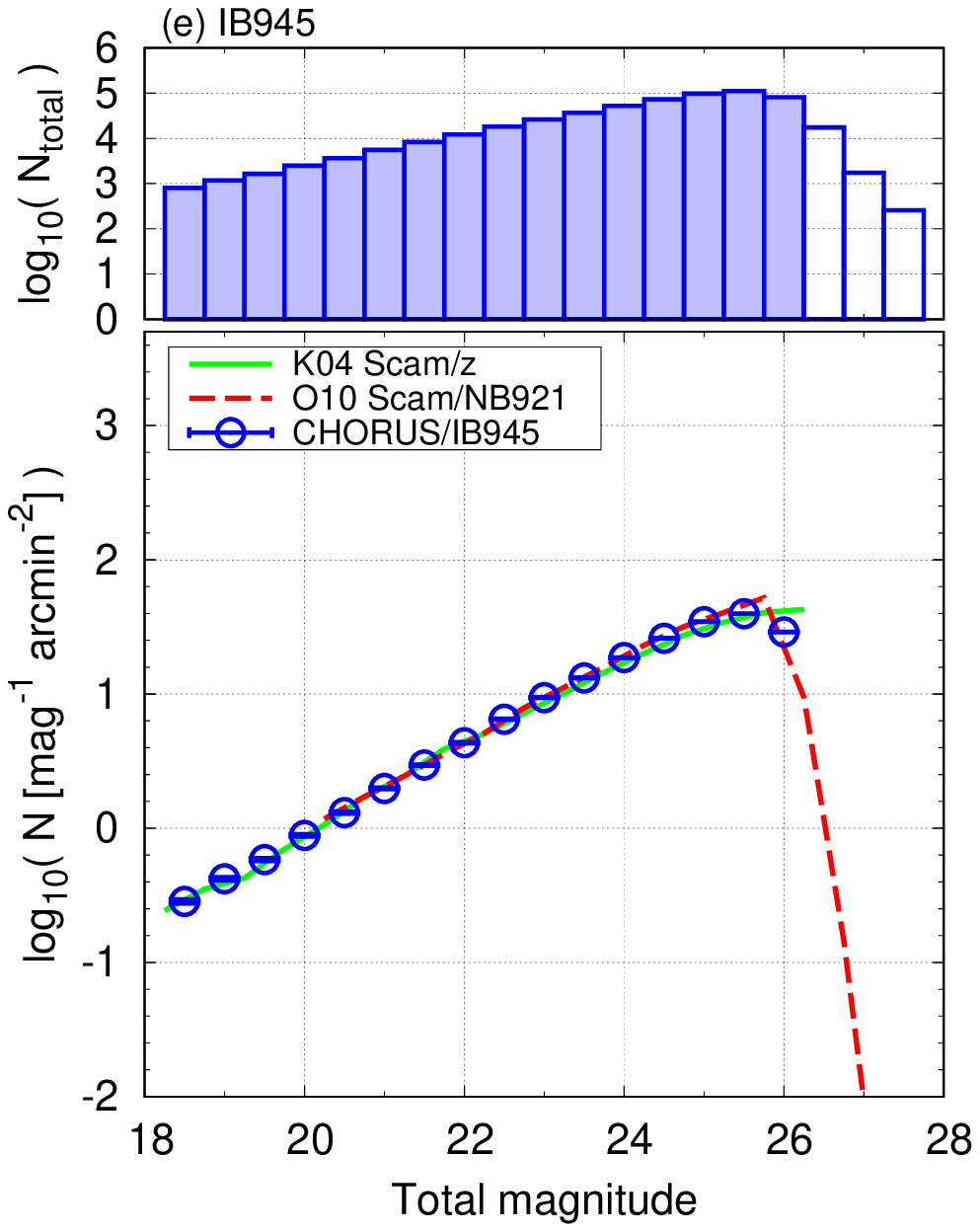}
  \end{center}
  \caption{Source number counts in NB387 (top left), NB527 (top middle), NB718 (top right), NB973 (bottom left) and IB945 (bottom right). For each band, upper panel shows histogram of actual numbers of detected sources and bottom panel shows number density per magnitude per square arcminutes. Empty histogram indicates magnitude range where detection completeness is less than $\sim 10 $\% and probability of spurious detections may be high. Number counts taken from literature are also shown in each panel for comparison. For NB387, we show both number counts with/without zero-point correction for catalog magnitudes produced by {\tt hscPipe} version 6.7. References of literature data: K04: \citet{Kashikawa2004}, O08: \citet{Ouchi2008}, O10: \citet{Ouchi2010}, Y12: \citet{Yamada2012}, N12: \citet{Nakajima2012}, S19: \citet{Sawicki2019}.} \label{fig:numbercount}
\end{figure*}

For the verification of the CHORUS PDR1 catalog, we calculated the source number counts in the CHORUS NBs and compared the results with those reported in the literature.
We adopted the {\tt undeblended{\_}convolvedflux} magnitudes ($0.84''$ FWHM convolution and $\phi 1.5''$ aperture case, except for NB387, where $1.1''$ FWHM convolution and $\phi 2.0''$ aperture case is adopted) as the total magnitude of the objects and excluded the objects and area in the masked regions.
The total magnitudes for each object were corrected for the Galactic extinction using the values listed in the CHORUS PDR1 catalog which are estimated from \citet{GalExtinc} dust maps (see also HSC SSP PDR papers \cite{Aihara2018b,Aihara2019}). 
These extinction magnitudes are small; for instance, $\sim 0.1$ mag even in NB387 in which the extinction effect is the severest.
We also applied a catalog flag, {\tt forced.merge{\_}peak{\_}XXXX} is {\tt True}, where {\tt XXXX} $=$ n387, n527, n718, i945, or n973, to ensure that we count only the sources detected in the NB that we are considering.
Otherwise, the number count included the sources that were not actually detected in the NB because {\tt hscPipe} lists all the sources detected in any single band in the catalogs.
Because the BB images of the HSC SSP are deeper than the CHORUS NB images, numerous faint BB sources, which are not detected in any of the CHORUS NBs, are also listed in the CHORUS PDR1 catalog.

Fig.~\ref{fig:numbercount} shows the number count measurements.
We only present the data points for those magnitudes at which the completeness (see Fig.~\ref{fig:completeness}) is larger than 10\%.
We also display the literature data of both the BBs and NBs whose wavelengths are close to those of each NB.
There are excellent agreements between all the CHORUS number counts and the literature data.
For example, the number counts of NB718, NB973, and IB945 present excellent agreement with the data of Subaru/Suprime-Cam ($i$ from \cite{Kashikawa2004} for NB718, and $z$ from \cite{Kashikawa2004} and NB921 from \cite{Ouchi2010} for NB973 and IB945).
The NB527 number count also agrees with those of Subaru/Suprim-Cam NB503 from \citet{Ouchi2008} and $V$ from \citet{Kashikawa2004}; however, there are slight differences from those in bluer bands, like $B$ from \citet{Kashikawa2004} and NB497 from \citet{Yamada2012}.
Because the slopes of the number counts reported in the literature are steeper at shorter wavelengths, the CHORUS results seem to follow this trend.
For NB387 after the zero-point correction (see sec.~\ref{sec:obs}), the number count data agree with those of Suprime-Cam/NB387 from \citet{Nakajima2012}, Suprime-Cam/$B$ from \citet{Kashikawa2004}, and CFHT/MegaCam/$u$ from \citet{Sawicki2019} at magnitudes fainter than 22.
At brighter magnitudes, the NB387 data of ours and Suprime-Cam lie between the two BB literature data, which is reasonable given the trend mentioned above, because the wavelength of NB387 is indeed between those of the $B$- and $u$-bands.
The zero-point uncorrected NB387 number count is less than those reported in the literature data. 
These results suggest the necessity and validity of the zero-point correction.

\section{How to use the catalog and example use cases of the CHORUS data}

The CHORUS PDR1 catalog is available at the Catalog Archive Server (CAS)\footnote{https://hsc-release.mtk.nao.ac.jp/datasearch/} of the HSC-SSP. 
An example SQL query to obtain a list of the coordinates, {\tt patch}, the Galactic extinction in NB387, $\phi 2.0''$ aperture NB387 magnitudes, NB387 total magnitudes ({\tt undeblended{\_}convolvedflux} with $1.1''$ FWHM convolution and $\phi 2.0''$ aperture), and the corresponding aperture corrections for the sources detected in NB387 is as follows:
\begin{description}
    \item {\tt SELECT}
    \begin{description}
        \item {\tt object{\_}id, ra, dec, patch, a{\_}n387}
        \item {\tt n387{\_}undeblended{\_}apertureflux{\_}20{\_}mag}
        \item {\tt n387{\_}undeblended{\_}convolvedflux{\_}2{\_}20{\_}mag}
        \item {\tt n387{\_}undeblended{\_}convolvedflux{\_}2{\_}20{\_}apcorr}
    \end{description}
    \item {\tt FROM}
    \begin{description}
        \item {\tt s18a{\_}chorus.forced}
        \item {\tt LEFT JOIN s18a{\_}chorus.forced3 USING (object{\_}id)}
        \item {\tt LEFT JOIN s18a{\_}chorus.forced6 USING (object{\_}id)}
    \end{description}
    \item {\tt WHERE}
    \begin{description}
        \item {\tt tractSearch(forced.object{\_}id, 9813)}
        \item {\tt AND forced.nchild = 0}
        \item {\tt AND forced.detect{\_}ispatchinner IS True}
        \item {\tt AND forced.merge{\_}peak{\_}n387 IS True}
        \item {\tt AND forced.n387{\_}pixelflags{\_}saturatedcenter IS False}
        \item {\tt AND forced.n387{\_}pixelflags{\_}bright{\_}object IS False}
    \end{description}
\end{description}
The CHORUS PDR1 ({\it forced}) catalog consists of 7 files of {\tt s18a{\_}chorus.forced}, {\tt s18a{\_}chorus.forced2}, ..., {\tt s18a{\_}chorus.forced7}.
The FITS files of the masked areas (Fig.~\ref{fig:mask}), spatial maps of PSF sizes (Fig.~\ref{fig:psf}) and limiting magnitudes (Fig.~\ref{fig:limitmag}), numerical data of detection completeness simulations (Fig.~\ref{fig:completeness}), and numerical data of measured number counts (Fig.~\ref{fig:numbercount}) are also available through the website\footnote{https://hsc-release.mtk.nao.ac.jp/doc/index.php/chorus/}.

In the following, we describe four specific studies ongoing with the CHORUS data and other ancillary studies including results published so far.

\subsection{Lyman continuum survey of galaxies and AGNs}

In the CHORUS project, we are performing a LyC survey for galaxies and AGNs at $z\simeq3.3$ and 4.9 using NB387 and NB527, respectively (see Fig.~\ref{fig:filters}). 
The samples are mainly the LAEs selected by NB527 ($z\simeq3.3$) and NB718 ($z\simeq4.9$). 
An {\it average} LyC-to-UV flux density ratio, or $f_{\rm esc}$, of the LAEs will be estimated by stacking analysis. 
Combining the LyC-to-UV ratios corrected for the IGM transmission and the observed UV luminosity densities, we can estimate the LyC luminosity densities and examine whether galaxies dominate the LyC emissivity at $z\sim3$--5. 
We can also detect the LyC from individual LAEs and AGNs. 
The number of such LyC galaxies at $z>3$ is still limited (e.g., \cite{Iwata2009,Vanzella2012,Vanzella2015,Steidel2018,Iwata2019}). 
This program will increase that number significantly.
Follow-up spectroscopy of these reliable LyC galaxies allows an $f_{\rm esc}$ calibration from the non-ionizing properties, such as a high [O~{\sc iii}]/[O~{\sc ii}] ratio \citep{Nakajima2014,deBarros2016,Nakajima2020}.
Once we establish the relation between the LyC emissivity (or $f_{\rm esc}$) values and the properties of galaxies observable at longer wavelengths at $z\lesssim5$, we can estimate $f_{\rm esc}$ of the galaxies in the epoch of reionization ($z>6$).
In this epoch, direct LyC observations are impossible owing to the severe absorption caused by the neutral hydrogen atoms in the IGM \citep{Inoue2008,Inoue2014}.

\subsection{Surveys of faint AGNs as dual-emitters}

Using NB718 and NB921, we can select a population of faint AGNs at $z\simeq4.9$ as Ly$\alpha$-C~{\sc iv} dual emitters if we appropriately treat the effects of broad lines whose line widths are broader than the NB widths (Fig.~\ref{fig:filters}). 
This selection method is unique because these faint AGNs with a magnitude range similar to that discussed by \citet{Giallongo2015} are difficult to be isolated from the LBGs by any combination of BB colors.
Moreover, our NB527 data allows the direct measurement of the LyC of these faint AGNs. 
The faint-AGN population reported by \citet{Giallongo2015} can reproduce the LyC emissivity required to keep the Universe ionized at $z\sim5$--6 if $f_{\rm esc}=1$ \citep{Madau2015}. 
We examine if this is the case not only by measuring the faint-AGN LF but also by measuring the $f_{\rm esc}$ of this population.
Simultaneously, we measure $f_{\rm esc}$ of galaxies at the same redshift and study their contribution directly. 
Subsequently, we identify the population that dominates the cosmic LyC emissivity at $z\sim3$--5 and infer the source of the cosmic reionization at $z>6$.

\subsection{Surveys of Pop~III-dominated galaxies}

Pop~III stars are expected to have high LyC emissivity and may play a significant role in the early phase of reionization (e.g., \cite{Sokasian2004}). 
However, observational evidence for this stellar population is still inconclusive (e.g., \cite{Kashikawa2012,Rydberg2015,Vanzella2020}), and its characteristics, e.g., the initial mass function and the LyC emissivity, are unknown.
Theoretically, Pop~III star clusters can be formed even at $z\sim2$--3, depending on the efficiency of the IGM metal-enrichment (e.g., \cite{Tornatore2007,Johnson2010,deSouza2011}). 
Indeed, pristine ($Z<10^{-4}$) gas clouds, from which Pop~III stars may form, are found at $z\simeq3$ (e.g., \cite{Fumagalli2011}). 
Identifying the epoch of the termination of a metal-free star formation is also a very important problem. 
The He~{\sc ii} emission line can be an indicator of Pop~III stars (e.g., \cite{Tumlinson2001,Schaerer2003}). 
Using our well-tuned set of NBs, we can search for Ly$\alpha$--He {\sc ii} dual emitters at $z\simeq2.2$, 3.3, and 4.9 (see Table~\ref{tab:filters} and Fig.~\ref{fig:filters}). 
Such a dual emitter search was performed by \citet{Nagao2008} with the Suprime-Cam for $z\sim4$. 
We update the upper limit on the He~{\sc ii} line luminosity density or the Pop~III star formation rate density by \citet{Nagao2008} with a 7 times larger area and up to $\sim1$ mag deeper imaging data at a wider redshift range of $z=2$--5. 
This will enable us to constrain the metal-enrichment efficiency in the IGM stronger than before.

\subsection{Mapping spatial distributions of ionized bubbles}

Because Ly$\alpha$ photons are sensitive to $x_{\rm HI}$, the distribution of the LAEs depends on both $x_{\rm HI}$ and the large-scale structure (LSS).
However, LBGs are selected by the continuum and are not very sensitive to $x_{\rm HI}$ but simply trace the LSS.
If we normalize the number density of the LAEs for each location by that of the LBGs, we can isolate the spatial distribution of $x_{\rm HI}$; 
we can visualize the $x_{\rm HI}$ map directly from the LAE-to-LBG density ratio. 
An essential point is to select LBGs with a similar redshift to the NB-selected LAEs whose redshift variation is as small as $\Delta z\simeq0.1$. 
Although such an LBG selection is not feasible by BB filters, we can reach $\Delta z\simeq0.3$ using an intermediate filter, IB945 for redshifts around $z=6.6$ and 7.0.
From the spatial variation of the observed number density ratio, {\it we can visualize an $x_{\rm HI}$ map far ahead of the future 21 cm experiments, such as the SKA.} 
Such an $x_{\rm HI}$ mapping opens a new pathway to examine the ionization topology, for example, allowing the measurements of the ionized bubble size directly.

\subsection{Ancillary sciences}

Very wide (1.6 deg$^2$) and very deep ($\gtrsim0.1L_*$) LAE samples at $z\simeq2.2$, 3.3, 4.9, 6.8, and 7.0 are obtained by this program. 
An LAE LF at $z=7.0$ has already been published as the CHORUS paper II \citep{Itoh2018} that reports a very similar LF shape to those at $z=5.7$ and 6.6 with a small overall decrement compared to the lower-$z$ ones.
This indicates that the reionization process does not alter the shape of the LAE LF in contrast to a previous claim of a bright-end hump \citep{Zheng2017}.
A comprehensive LAE search in the CHORUS data in conjunction with the HSC SSP data is  ongoing, adopting a machine learning technique (Ono, Y. et al.\ in preparation).
A study of intensity mapping of Ly$\alpha$ emission is also ongoing (Kikuchihara, S. et al.\ in preparation).
More LAE related studies including environmental ones and optical emission line properties in the near-infrared band are also ongoing.
We are also searching for spatially extended LAEs --- Ly$\alpha$ blobs --- in the imaging data. 
A result of a Ly$\alpha$ blob survey at $4.9\leq z \leq 7.0$ has been published recently as the CHORUS paper III \citep{Zhang2020} that reports a Ly$\alpha$ blob at $z\simeq7.0$, the highest-$z$ blob found so far.
[O {\sc ii}], [O {\sc iii}], H$\beta$, and H$\alpha$ emitters at various redshifts are also studied with the CHORUS data in conjunction with the HSC SSP data \citep{Hayashi2020} and Spitzer data \citep{Harikane2018}.
More line emitter studies including AGN surveys will be performed in future.
The CHORUS PDR1 data will be useful for various studies performed all over the world.
We strongly encourage the readers' own studies using the dataset.

\section{Summary}

In this study, we conducted a deep imaging survey using a set of narrow-band (NB) and intermediate-band (IB) filters, named as Cosmic HydrOgen Reionization Unveiled with Subaru (CHORUS), with the Hyper Suprime-Cam (HSC) equipped on the Subaru Telescope.
The filter set used in the survey is shown in Figure~\ref{fig:filters}.
The wavelengths, NB387, NB527, NB718, NB973, and IB945, summarized in Table~\ref{tab:filters}, were chosen in a very well-organized manner to maximize the scientific outcomes when the filters were used in combination with another NB, NB921, from the Subaru Strategic Program for HSC (HSC SSP).
The observations were performed on 18 nights during the period from January 2017 to December 2018 in an intensive program of the Subaru Telescope.
The observing field was the COSMOS field \citep{Scoville2007}.
The data reduction, source detection, and multiband photometry were conducted using {\tt hscPipe}, the official pipeline software for the HSC developed by the HSC SSP team \citep{Bosch2018}.
The {\tt hscPipe} version used was 6.7, the same as that used for HSC SSP public data release (PDR) 2 \citep{Aihara2019}.
Because there is a known problem for the {\tt cmodel} magnitudes in this version, we recommended using the {\tt undeblended{\_}convolvedflux} magnitudes instead.
The imaging data qualities were examined thoroughly in a spatially dependent manner, as shown in Figures~\ref{fig:psf} and \ref{fig:limitmag}.
For each NB/IB, typical values of the size of point spread functions, $5\sigma$ limiting magnitude, and survey area excluding the carefully defined masked regions (see Figure~\ref{fig:mask} as an example) are summarized in Table~\ref{tab:obs}.
An extensive set of mock observation simulations was performed to estimate the source detection completeness, as shown in Figure~\ref{fig:completeness}.
We checked the source number counts and compared them with previous studies in Figure~\ref{fig:numbercount}.
An excellent agreement with the literature data was obtained in each NB/IB.
All the images and photometric catalogs are publicly available through the HSC SSP database.
This data release is called CHORUS PDR 1.

\begin{ack}
The CHORUS NB/IB filter development was supported by JSPS KAKENHI Grant Numbers 23244033 (KS: NB387), 24244018 (II: NB527), 23244025 (MO: NB921 and NB973), and 23684010 (AKI: IB945), and by a special operating cost grant by MEXT to Ehime University (YT: NB718).
This work was also supported by JSPS KAKENHI Grant Numbers 26287034 (AKI, KM), 17H01114 (AKI, KM, SY, MO, II, KS, TN, NK, YO), 15H02064 (MO), and 17H01110 (MO).

The authors would like to thank the referee for constructive comments which were useful to improve the quality of the manuscript.
The authors would like to acknowledge their specific contributions by the following people: 
Ryohei Itoh and Ryota Kakuma for data reduction, 
Satoshi Kawanomoto and Masakazu A.\ R.\ Kobayashi for filter development and inspection, 
Karin Shimodate for the observations, and 
Masayuki Tanaka for public data release through the HSC SSP database.
The authors would also like to thank the following members of the CHORUS project for their various contributions: 
Naoki Yasuda, 
Masayuki Umemura, 
Masao Mori, 
Yasuhito Shioya,
Toru Yamada, 
Ikkoh Shimizu, 
Kenji Hasegawa, 
Tomoaki Ishiyama, 
Mana Niida, 
Akira Konno, 
Shiro Mukae, 
Andrea Schulze, 
Genoveva Micheva, 
Masayuki Akiyama, 
Kenichi Tadaki, 
Michael Strauss, 
Ryo Higuchi, 
Takashi Kojima, 
Masafusa Onoue, 
Yoshiki Matsuoka, 
Masatoshi Imanishi, 
Rhythm Shimakawa, 
Takuya Hashimoto, 
Y.-T. Lin, 
John Silverman, 
Seiji Fujimoto, 
Kohei Iwashita, 
Takuji Yamashita, 
Tomoko Suzuki, and 
Hisakazu Uchiyama

The Hyper Suprime-Cam (HSC) collaboration includes the astronomical communities of Japan and Taiwan, and Princeton University. The HSC instrumentation and software were developed by the National Astronomical Observatory of Japan (NAOJ), the Kavli Institute for the Physics and Mathematics of the Universe (Kavli IPMU), the University of Tokyo, the High Energy Accelerator Research Organization (KEK), the Academia Sinica Institute for Astronomy and Astrophysics in Taiwan (ASIAA), and Princeton University.  Funding was contributed by the FIRST program from the Japanese Cabinet Office, the Ministry of Education, Culture, Sports, Science and Technology (MEXT), the Japan Society for the Promotion of Science (JSPS), Japan Science and Technology Agency  (JST), the Toray Science  Foundation, NAOJ, Kavli IPMU, KEK, ASIAA, and Princeton University.

This paper makes use of software developed for the Large Synoptic Survey Telescope. We thank the LSST Project for making their code available as free software at  http://dm.lsst.org

This paper is based on data collected at the Subaru Telescope and retrieved from the HSC data archive system, which is operated by Subaru Telescope and Astronomy Data Center (ADC) at NAOJ. Data analysis was in part carried out with the cooperation of Center for Computational Astrophysics (CfCA), NAOJ.

The Pan-STARRS1 Surveys (PS1) and the PS1 public science archive have been made possible through contributions by the Institute for Astronomy, the University of Hawaii, the Pan-STARRS Project Office, the Max Planck Society and its participating institutes, the Max Planck Institute for Astronomy, Heidelberg, and the Max Planck Institute for Extraterrestrial Physics, Garching, The Johns Hopkins University, Durham University, the University of Edinburgh, the Queen’s University Belfast, the Harvard-Smithsonian Center for Astrophysics, the Las Cumbres Observatory Global Telescope Network Incorporated, the National Central University of Taiwan, the Space Telescope Science Institute, the National Aeronautics and Space Administration under grant No. NNX08AR22G issued through the Planetary Science Division of the NASA Science Mission Directorate, the National Science Foundation grant No. AST-1238877, the University of Maryland, Eotvos Lorand University (ELTE), the Los Alamos National Laboratory, and the Gordon and Betty Moore Foundation.

The authors would like to thank Editage (www.editage.com) for English language editing.

\end{ack}




\begin{thebibliography}{}
\bibitem[Aihara et al.(2018a)]{Aihara2018a}
Aihara, H., Arimoto, N., Armstrong, R., Arnouts, S., Bahcall, N. A., 
Bickerton, S., Bosch, J., Bundy, K., et al.\
2018a, PASJ, 70, S4

\bibitem[Aihara et al.(2018b)]{Aihara2018b}
Aihara, H., Armstrong, R., Bickerton, S., Bosch, J., Coupon, J., 
Furusawa, H., Hayashi, Y., Ikeda, H., et al.\
2018b, PASJ, 70, S8

\bibitem[Aihara et al.(2019)]{Aihara2019}
Aihara, H., AlSayyad, Y., Ando, M., Armstrong, R., Bosch, J., 
Egami, E., Furusawa, H., Furusawa, J., et al.\
2019, PASJ, 71, 114

\bibitem[Bagley et al.(2017)]{Bagley2017}
Bagley, M. B., Scarlata, C., Henry, A., Rafelski, M., Malkan, M., 
Teplitz, H., Dai, Y. S., Baronchelli, I., et al.\
2017, ApJ, 837, 11

\bibitem[Becker et al.(2007)]{Becker2007}
Becker, G. D., Rauch, M., Sargent, W. L. W.\
2007, ApJ, 662, 72

\bibitem[Becker et al.(2015)]{Becker2015}
Becker, G. D., Bolton, J. S., Madau, P., Pettini, M., Ryan-Weber, E. V., Venemans, B. P.\
2015, MNRAS, 447, 3402

\bibitem[Bertin \& Arnouts(1996)]{sextractor}
Bertin, E., \& Arnouts, S.\ 1996, A\&AS, 117, 393

\bibitem[Bianchi et al.(2001)]{Bianchi2001}
Bianchi, S., Cristiani, S., Kim, T.-S.\ 
2001, A\&A, 376, 1

\bibitem[Bosch et al.(2018)]{Bosch2018}
Bosch, J., Armstrong, R., Bickerton, S., Furusawa, H., Ikeda, H., Koike, M., 
Lupton, R., Mineo, S., et al.\
2018, PASJ, 70, S5

\bibitem[Carlberg(1981)]{Carlberg1981}
Carlberg, R. G.\ 1981, MNRAS, 197, 1021

\bibitem[Castellano et al.(2016)]{Castellano2016}
Castellano, M., Dayal, P., Pentericci, L., Fontana, A., Hutter, A., 
Brammer, G., Merlin, E., Grazian, A., et al.\
2016, ApJ, 818, L3

\bibitem[Chambers et al.(2016)]{Chambers2016}
Chambers, K. C., Magnier, E. A., Metcalfe, N., Flewelling, H. A., 
Huber, M. E., Waters, C. Z., Denneau, L., Draper, P. W., et al.\
2016, arXiv:1612.05560

\bibitem[de Barros et al.(2016)]{deBarros2016}
de Barros, S., Vanzella, E., Amor{\'i}n, R., Castellano, M., Siana, B., 
Grazian, A., Suh, H., Balestra, I., et al.\
2016, A\&A, 585, 51

\bibitem[de Souza et al.(2011)]{deSouza2011}
de Souza, R. S., Yoshida, N., Ioka, K.\ 2011, A\&A, 533, 32

\bibitem[Fan et al.(2006)]{Fan2006}
Fan, X., Strauss, M. A., Becker, R. H., White, R. L., Gunn, J. E., 
Knapp, G. R., Richards, G. T., Schneider, D. P., et al.\
2006, AJ, 132, 117

\bibitem[Finkelstein et al.(2019)]{Finkelstein2019}
Finkelstein, S. L., D'Aloisio, A., Paardekooper, J.-P., Ryan, R. Jr., Behroozi, P., 
Finlator, K., Livermore, R., Upton Sanderbeck, P. R., et al.\
2019, ApJ, 879, 36

\bibitem[Fumagalli et al.(2011)]{Fumagalli2011}
Fumagalli, M., O'Meara, J. M., Prochaska, J. X.\
2011, Science, 334, 1245

\bibitem[Furusawa et al.(2018)]{Furusawa2018}
Furusawa, H., Koike, M., Takata, T., Okura, Y., Miyatake, H., 
Lupton, R. H., Bickerton, S., Price, P. A., et al.\
2018, PASJ, 70, S3

\bibitem[Giallongo et al.(2015)]{Giallongo2015}
Giallongo, E., Grazian, A., Fiore, F., Fontana, A., Pentericci, L., 
Vanzella, E., Dickinson, M., Kocevski, D., et al.\
2015, A\&A, 578, A83

\bibitem[Grazian et al.(2018)]{Grazian2018}
Grazian, A., Giallongo, E., Boutsia, K., Cristiani, S., Vanzella, E., 
Scarlata, C., Santini, P., Pentericci, L., et al.\
2018, A\&A, 613, A44

\bibitem[Greig \& Mesinger(2017)]{GreigMesinger17}
Greig, B., \& Mesinger, A.\ 2017, MNRAS, 465, 4838

\bibitem[Harikane et al.(2018)]{Harikane2018}
Harikane, Y., Ouchi, M., Shibuya, T., Kojima, T., Zhang, H., Itoh, R., 
Ono, Y., Higuchi, R., et al.\
2018, ApJ, 859, 84

\bibitem[Harikane et al.(2019)]{Harikane2019}
Harikane, Y., Ouchi, M., Ono, Y., Fujimoto, S., Donevski, D., Shibuya, T., 
Faisst, A. L., Goto, T., et al.\
2019, ApJ, 883, 142

\bibitem[Hasegawa et~al.(2016)]{Hasegawa2016}
Hasegwa, K., Asaba, S., Ichiki, K., Inoue, A. K., 
Inoue, S., Ishiyama, T., Shimabukuro, H., Takahashi, K., et~al.\ 
2016, arXiv:1603.01961

\bibitem[Hayashi et al.(2020)]{Hayashi2020}
Hayashi, M., Shimakawa, R., Tanaka, M., Onodera, M., Koyama, Y., Inoue, A. K., 
Komiyama, Y., Lee, C.-H., et al.\ 
2020, PASJ, submitted

\bibitem[Higuchi et al.(2019)]{Higuchi2019}
Higuchi, R., Ouchi, M., Ono, Y., Shibuya, T., Toshikawa, J., Harikane, Y., 
Kojima, T., Chiang, Y.-K., et al.\
2019, ApJ, 879, 28

\bibitem[Hu et al.(2010)]{Hu2010}
Hu, E. M., Cowie, L. L., Barger, A. J., Capak, P., Kakazu, Y., Trouille, L.\
2010, ApJ, 725, 394

\bibitem[Hu et al.(2019)]{Hu2019}
Hu, W., Wang, J., Zheng, Z.-Y., Malhotra, S., Rhoads, J. E., Infante, L., 
Barrientos, L. F., Yang, H., et al.\
2019, ApJ, 886, 90

\bibitem[Iliev et al.(2006)]{Iliev2006}
Iliev, I. T., Mellema, G., Pen, U.-L., Merz, H., Shapiro, P. R., 
Alvarez, M. A.\ 2006, MNRAS, 369, 1625

\bibitem[Inoue et al.(2006)]{Inoue2006}
Inoue, A. K., Iwata, I., Deharveng, J.-M.\ 
2006, MNRAS, 371, L1

\bibitem[Inoue \& Iwata(2008)]{Inoue2008}
Inoue, A. K., Iwata, I.\ 2008, MNRAS, 387, 1681

\bibitem[Inoue et al.(2014)]{Inoue2014}
Inoue, A. K., Shimizu, I., Iwata, I., Tanaka, M.\
2014, MNRAS, 442, 1805

\bibitem[Inoue et al.(2018)]{Inoue2018}
Inoue, A. K., Hasegawa, K., Ishiyama, T., Yajima, H., Shimizu, I., Umemura, M., 
Konno, A., Harikane, Y., et al.\
2018, PASJ, 70, 55

\bibitem[Itoh et al.(2018)]{Itoh2018}
Itoh, R., Ouchi, M., Zhang, H., Inoue, A. K., Mawatari, K., Shibuya, T., 
Harikane, Y., Ono, Y., et al.\ 
2018, ApJ, 867, 46

\bibitem[Iwata et al.(2009)]{Iwata2009}
Iwata, I., Inoue, A. K., Matsuda, Y., Furusawa, H., Hayashino, T., Kousai, K., 
Akiyama, M., Yamanda, T., Burgarella, D., Deharveng, J.-M.\
2009, ApJ, 692, 1287

\bibitem[Iwata et al.(2019)]{Iwata2019}
Iwata, I., Inoue, A. K., Micheva, G., Matsuda, Y., Yamada, T.\
2019, MNRAS, 488, 5671

\bibitem[Johnson(2010)]{Johnson2010}
Johnson, J. L.\ 2010, MNRAS, 404, 1425

\bibitem[Kashikawa et al.(2004)]{Kashikawa2004}
Kashikawa, N., Shimasaku, K., Yasuda, N., Ajiki, M., Akiyama, M., Ando, H., 
Aoki, K., Doi, M., et al.\
2004, PASJ, 56, 1011

\bibitem[Kashikawa et al.(2006)]{Kashikawa2006}
Kashikawa, N., Shimasaku, K., Malkan, M. A., Doi, M., Matsuda, Y., 
Ouchi, M., Taniguchi, Y., Ly, C., et al.\
2006, ApJ, 648, 7

\bibitem[Kashikawa et al.(2011)]{Kashikawa2011}
Kashikawa, N., Shimasaku, K., Matsuda, Y., Egami, E., Jiang, L., 
Nagao, T., Ouchi, M., Malkan, M. A., et al.\
2011, ApJ, 734, 119

\bibitem[Kashikawa et al.(2012)]{Kashikawa2012}
Kashikawa, N., Nagao, T., Toshikawa, J., Ishizaki, Y., Egami, E., Hayashi, M., 
Ly, C., Malkan, M. A., et al.\ 
2012, ApJ, 761, 85

\bibitem[Kashikawa et al.(2015)]{Kashikawa2015}
Kashikawa, N., Ishizaki, Y., Willott, C. J., Onoue, M., Im, M., 
Furusawa, H., Toshikawa, J., Ishikawa, S., et al.\
2015, ApJ, 798, 28

\bibitem[Kawamata et al.(2018)]{Kawamata2018}
Kawamata, R., Ishigaki, M., Shimasaku, K., Oguri, M., Ouchi, M., Tanigawa, S.\
2018, ApJ, 855, 4

\bibitem[Kawanomoto et al.(2018)]{Kawanomoto2018}
Kawanomoto, S., Uraguchi, F., Komiyama, Y., Miyazaki, S., Furusawa, H., 
Finet, F., Hattori, T., Wang, S.-Y., et al.\
2018, PASJ, 70, 66

\bibitem[Komatsu et al.(2011)]{Komatsu2011}
Komatsu, E., Smith, K. M., Dunkley, J., Bennett, C. L., Gold, B., 
Hinshaw, G., Jarosik, N., Larson, D., et al.\
2011, ApJS, 192, 18

\bibitem[Komiyama et al.(2018)]{Komiyama2018}
Komiyama, Y., Obuchi, Y., Nakaya, H., Kamata, Y., Kawanomoto, S., 
Utsumi, Y., Miyazaki, S., Uraguchi, F., et al.\
2018, PASJ, 70, S2

\bibitem[Konno et al.(2014)]{Konno2014}
Konno, A., Ouchi, M., Ono, Y., Shimasaku, K., Shibuya, T., Furusawa, H., 
Nakajima, K., Naito, Y., et al.\ 2014, ApJ 797, 16

\bibitem[Konno et al.(2018)]{Konno2018}
Konno, A., Ouchi, M., Shibuya, T., Ono, Y., Shimasaku, K., 
Taniguchi, Y., Nagao, T., Kobayashi, M. A. R., et al.\ 
2018, PASJ, 70, 16

\bibitem[Liang et al.(2020)]{Liang2020}
Liang, Y., Kashikawa, N., Cai, Z., Fan, X., Prochaska, J. X., Shimasaku, K., 
Tanaka, M., Uchiyama, H., et al.\
2020, ApJ, in press (arXiv:2008.01733)

\bibitem[Madau \& Haardt(2015)]{Madau2015}
Madau, P., Haardt, F.\ 2015, ApJ, 813, L8

\bibitem[Matsuoka et al.(2018)]{Matsuoka2018}
Matsuoka, Y., Strauss, M. A., Kashikawa, N., Onoue, M., Iwasawa, K., Tang, J.-J., 
Lee, C.-H., Imanishi, M., et al.\
2018, ApJ, 869, 150

\bibitem[Matthee et al.(2015)]{Matthee2015}
Matthee, J., Sobral, D., Santos, S., R{\''o}ttgering, H., Darvish, B., Mobasher, B.\
2015, MNRAS, 451, 400

\bibitem[Micheva et al.(2017)]{Micheva2017}
Micheva, G., Iwata, I., Inoue, A. K.,\
2017, MNRAS, 465, 302

\bibitem[Miralda-Escud{\'e} et al.(2000)]{Miralda-Escude2000}
Miralda-Escud{\'e}, J., Haehnelt, M., Rees, M. J.\ 
2000, ApJ, 530, 1

\bibitem[Miyazaki et al.(2018)]{Miyazaki2018}
Miyazaki, S., Komiyama, Y., Kawanomoto, S., Doi, Y., Furusawa, H., 
Hamana, T., Hayashi, Y., Ikeda, H., et al.\
2018, PASJ, 70, S1

\bibitem[Nagao et al.(2005)]{Nagao2005}
Nagao, T., Motohara, K., Maiolino, R., Marconi, R., Taniguchi, Y., 
Aoki, K., Ajiki, M., Shioya, Y.\
2005, ApJ, 631, L5

\bibitem[Nagao et al.(2008)]{Nagao2008}
Nagao, T., Sasaki, S. S., Maiolino, R., Grady, C., Kashikawa, N., 
Ly, C., Malkan, M. A., Motohara, K., et al.\
2008, ApJ, 680, 100

\bibitem[Nakajima et al.(2012)]{Nakajima2012}
Nakajima, K., Ouchi, M., Shimasaku, K., Ono, Y., Lee, J. C., 
Foucaud, S., Ly, C., Dale, D. A., et al.\
2012, ApJ, 745, 12

\bibitem[Nakajima \& Ouchi(2014)]{Nakajima2014}
Nakajima, K., Ouchi, M.\ 2014, MNRAS, 442, 900

\bibitem[Nakajima et al.(2020)]{Nakajima2020}
Nakajima, K., Ellis, R. S., Robertson, B. E., Tang, M., Stark, D. P.\
2020, ApJ, 889, 161

\bibitem[Naidu et al.(2020)]{Naidu2020}
Naidu, R. P., Tacchella, S., Mason, C. A., Bose, S., Oesch, P. A., Conroy, C.\ 
2020, ApJ, in press (arXiv:1907.13130)

\bibitem[Nakamura et al.(2011)]{Nakamura2011}
Nakamura, E., Inoue, A. K., Hayashino, T., Horie, M., Kousai, K., Fujii, T., Matsuda, Y.\
2011, MNRAS, 412, 2579

\bibitem[Oke (1990)]{Oke1990}
Oke, J. B., 1990, AJ, 99, 1621

\bibitem[Onoue et al.(2017)]{Onoue2017}
Onoue, M., Kashikawa, N., Willot, C. J., Hibon, P., Im, M., Furusawa, H., 
Harikane, Y., Imanishi, M., et al.\
2017, ApJ, 847, L15

\bibitem[Ouchi et al.(2008)]{Ouchi2008}
Ouchi, M., Shimasaku, K., Akiyama, M., Simpson, C., Saito, T., 
Ueda, Y., Furusawa, H., Sekiguchi, K., et al.\
2008, ApJS, 176, 301

\bibitem[Ouchi et al.(2010)]{Ouchi2010}
Ouchi, M., Shimasaku, K., Furusawa, H., Saito, T., Yoshida, M., 
Akiyama, M., Ono, Y., Yamada, T., et al.\ 2010, ApJ, 723, 869

\bibitem[Ouchi et al.(2018)]{Ouchi2018}
Ouchi, M., Harikane, Y., Shibuya, T., Shimasaku, K., Taniguchi, Y., 
Konno, A., Kobayashi, M. A. R., Kajisawa, M., et al.\ 
2018, PASJ, 70, S13

\bibitem[Parsa et al.(2018)]{Parsa2018}
Parsa, S., Dunlop, J. S., McLure, R. J.\
2018, MNRAS, 474, 2904

\bibitem[Pickles(1998)]{Pickles1998}
Pickles, A. J., 1998, PASP, 110, 863

\bibitem[Planck collaboration(2018)]{Planck2018}
Planck collaboration 2018, A\&A, submitted (arXiv:1807.06209)

\bibitem[Robertson et al.(2015)]{Robertson2015}
Robertson, B. E., Ellis, R. S., Furlanetto, S. R., Dunlop, J. S.\
2015, ApJ, 802, L19

\bibitem[Rowe et al.(2015)]{GALSIM}
Rowe, B.~T.~P., Jarvis, M., Mandelbaum, R., et al.\
2015, Astronomy and Computing, 10, 121

\bibitem[Rydberg et al.(2015)]{Rydberg2015}
Rydberg, C.-E., Zackrisson, E., Zitrin, A., Guaita, L., 
Melinder, J., Asadi, S., Gonzalez, J., {\''O}stlin, G., Str{\''o}m, T.\
2015, ApJ, 804, 13

\bibitem[Santos et al.(2016)]{Santos2016}
Santos, S., Sobral, D., Matthee, J.\
2016, MNRAS, 463, 1678

\bibitem[Sawicki et al.(2019)]{Sawicki2019}
Sawicki, M., Arnouts, S., Huang, J., Coupon, J., Golob, A., Gwyn, S., 
Foucaud, S., Moutard, T., et al.\
2019, MNRAS, 489, 5202

\bibitem[Schaerer(2003)]{Schaerer2003}
Schaerer, D.\  2003, A\&A, 397, 527

\bibitem[Scoville et al.(2007)]{Scoville2007}
Scoville, N., Aussel, H., Brusa, M., Capak, P., Carollo, C. M., 
Elvis, M., Giavalisco, M., Guzzo, L., et al.\
2007, ApJS, 171, 1

\bibitem[Shibuya et al.(2015)]{Shibuya2015}
Shibuya, T., Ouchi, M., Harikane, Y.\
2015, ApJS, 219, 15

\bibitem[Shibuya et al.(2018a)]{Shibuya2018a}
Shibuya, T., Ouchi, M., Konno, A., Higuchi, R., Harikane, Y., Ono, Y.,
Shimasaku, K., Taniguchi, Y., et al.\ 
2018a, PASJ, 70, S14

\bibitem[Shibuya et al.(2018b)]{Shibuya2018b}
Shibuya, T., Ouchi, M., Harikane, Y., Rauch, M., Ono, Y., Mukae, S., 
Higuchi, R., Kojima, T., et al.\ 
2018b, PASJ, 70, S15

\bibitem[Schlegel et al.(1998)]{GalExtinc}
Schlegel, D.~J., Finkbeiner, D.~P., \& Davis, M.\
1998, \apj, 500, 525

\bibitem[Steidel et al.(2018)]{Steidel2018}
Steidel, C. C., Bogosavljevi{\'c}, M., Shapley, A. E., Reddy, N. A., 
Rudie, G. C., Pettini, M., Trainor, R. F., Strom, A. L.\
2018, ApJ, 869, 123

\bibitem[Sokasian et al.(2004)]{Sokasian2004}
Sokasian, A., Yoshida, N., Abel, T., Hernquist, L., Springel, V.\
2004, MNRAS, 350, 47

\bibitem[Taylor et al.(2020)]{Taylor2020}
Taylor, A. J., Barger, A. J., Cowie, L. L., Hu, E. M., Songaila, A.\
2020, ApJ, in press (arXiv:2004.09510)

\bibitem[Tilvi et al.(2020)]{Tilvi2020}
Tilvi, V., Malhotra, S., Rhoads, J. E., Coughlin, A., Zheng, Z., 
Finkelstein, S. L., Veilleux, S., Mobasher, B., et al.\
2020, ApJ, 891, L10

\bibitem[Tornatore et al.(2007)]{Tornatore2007}
Tornatore, L., Ferrara, A., Schneider, R.\ 2007, MNRAS, 382, 945

\bibitem[Tumlinson et al.(2001)]{Tumlinson2001}
Tumlinson, J., Giroux, M. L., Shull, J. M.\
2001, ApJ, 550, L1

\bibitem[Vanzella et al.(2012)]{Vanzella2012}
Vanzella, E., Guo, Y., Giavalisco, M., Grazian, A., Castellano, M., 
Cristiani, S., Dickinson, M., Fontana, A., et al.\
2012, ApJ, 751, 70

\bibitem[Vanzella et al.(2015)]{Vanzella2015}
Vanzella, E., de Barros, S., Castellano, M., Grazian, A., Inoue, A. K., 
Schaerer, D., Guaita, L., Zamorani, G., et al.\
2015, A\&A, 576, 116

\bibitem[Vanzella et al.(2020)]{Vanzella2020}
Vanzella, E., Meneghetti, M., Caminha, G. B., Castellano, M., Calura, F., 
Rosati, P., Grillo, C., Dijkstra, M., et al.\
2020, MNRAS, 494, L81

\bibitem[Yamada et al.(2012)]{Yamada2012}
Yamada, T., Nakamura, Y., Matsuda, Y., Hayashino, T., Yamauchi, R., 
Morimoto, N., Kousai, K., Umemura, M.\ 2012, AJ, 143, 79

\bibitem[Zhang et al.(2020)]{Zhang2020}
Zhang, H., Ouchi, M., Itoh, R., Shibuya, T., Ono, Y., Harikane, Y., 
Inoue, A. K., Rauch, M., et al.\
2020, ApJ, 891, 177

\bibitem[Zheng et al.(2017)]{Zheng2017}
Zheng, Z.-Y., Wang, J., Rhoads, J., Infante, L., Malhotra, S., Hu, W., 
Walker, A. R., Jiang, L., et al.\
2017, ApJ, 842, L22

\end{thebibliography}
\end{document}